\documentclass[prl, onecolumn, amsmath, amssymb,superscriptaddress]{revtex4-2}

\usepackage[none]{hyphenat}
\usepackage{amsmath}
\usepackage{amssymb}
\usepackage{graphicx}
\usepackage{grffile}
\usepackage{bbold}
\graphicspath{{images/}}
\graphicspath{{Figures/}}

\usepackage{titlesec}

\usepackage{tikz}
\usepackage{mysetup}

\usepackage{calc}
\usepackage{cancel}
\usepackage{xstring}
\usepackage{mathtools}
\usepackage{color}
\usepackage{cancel}
\usepackage[centering,hmargin=2.0cm]{geometry}

\usepackage{braket}
\usepackage{bbm}

\newcommand{\K}{\textnormal{K}}
\newcommand{\ci}{\textnormal{C}}
\newcommand{\li}{\textnormal{L}}
\newcommand{\ri}{\textnormal{R}}
\newcommand{\lli}{\tilde{\textnormal{L}}}
\newcommand{\rri}{\tilde{\textnormal{R}}}
\newcommand{\si}{\textnormal{S}}
\newcommand{\tn}{\textnormal}

\newcommand{\calrL}{{\cal G}^{\textnormal{ret},\cancel{\li}}}
\newcommand{\calrR}{{\cal G}^{\textnormal{ret},\cancel{\ri}}}
\newcommand{\calrLR}{{\cal G}^{\textnormal{ret},\cancel{\li}\cancel{\ri}}}

\newcommand{\calaL}{{\cal G}^{\textnormal{adv},\cancel{\li}}}
\newcommand{\calaR}{{\cal G}^{\textnormal{adv},\cancel{\ri}}}
\newcommand{\calaLR}{{\cal G}^{\textnormal{adv},\cancel{\li}\cancel{\ri}}}

\newcommand{\calkLR}{{\cal G}^{\textnormal{K},\cancel{\li}\cancel{\ri}}}

\newcommand{\tsk}{\tilde\Sigma^{\tn{K}}}

\newcommand{\calSrR}{{\cal S}^{\textnormal{ret},\cancel{\ri}}}
\newcommand{\calSrLR}{{\cal S}^{\textnormal{ret},\cancel{\li}\cancel{\ri}}}

\newcommand{\calSaR}{{\cal S}^{\textnormal{adv},\cancel{\ri}}}
\newcommand{\calSaLR}{{\cal S}^{\textnormal{adv},\cancel{\li}\cancel{\ri}}}

\newcommand{\calSkLR}{{\cal S}^{\textnormal{K},\cancel{\li}\cancel{\ri}}}

\newcommand{\pa}{\partial_\Lambda^*}
\newcommand{\la}{\Lambda}

\makeatletter
\renewcommand{\p@subsection}{}
\renewcommand{\p@subsubsection}{}
\renewcommand{\p@paragraph}{}
\makeatother

\newcommand{\syst}[2]{
\begin{tikzpicture}
    \def\efi{#1}
    \coordinate(A0) at (0.6cm,0cm);
    \foreach \a in {1,2,...,7}{
        \if\a1%
            \node[circle, minimum size=0.6cm, below right=\efi*\a and 1.2cm*\a of A0, thick](A\a){$\dots$};
        \else
            \if\a7%
                \node[circle, minimum size=0.6cm, below right=\efi*\a and 1.2cm*\a of A0, thick](A\a){$\dots$};
            \else
                \node[draw, circle, minimum size=0.6cm, below right= \efi*\a and 1.2cm*\a of A0, thick](A\a){};
            \fi
        \fi;
    }
    \foreach[evaluate=\a as \an using int(\a+1)] \a in {1,2,...,6}
        \path[]
            (A\a.north east) edge [
              text=black,
              thick,
              shorten <=2pt,
              shorten >=2pt,
              bend left=50
              ] node[above]{\IfEqCase{#2}{%
                              {1}{\(t\)}%
                              {0}{\(\)}%
                             }%
                           } (A\an.north west);
    \foreach[evaluate=\a as \an using int(\a+1)] \a in {1,2,...,6}
        \path[]
            (A\a.south east) edge [
              text=black,
              thick,
              shorten <=2pt,
              shorten >=2pt,
              bend right=50,
              dashed
              ] node[below]{\IfEqCase{#2}{%
                              {1}{\(U\)}%
                              {0}{\(\)}%
                             }%
                           } (A\an.south west);

    \foreach \a in {2,3,...,6}{
          \coordinate[below=2.6cm of A\a](r\a);
          \draw[thick, shading=axis, bottom color=white, top color=custBlue] (r\a)++(-0:0.4 and 1.5) arc(-0:180:0.4 and 1.5);
        \path[]
            (r\a)++(90:0.4 and 1.5) edge [
              text=black,
              thick,
              shorten <=2pt,
              shorten >=2pt,
              bend right=50
              ] node[left]{\IfEqCase{#2}{%
                              {1}{\(\Gamma\)}%
                              {0}{\(\)}%
                             }%
                           } (A\a.south);
    }
    \node[above=1.5cm of A2] (ar){};
    \draw[->] (ar) --node[above]{$E$} +(1.2cm,-\efi);
\end{tikzpicture}
}

\begin{document}
\title{Review of recent developments of the functional renormalization group for systems out of equilibrium}

\author{G.~Camacho}
\affiliation{Technische Universit\"at Braunschweig, Institut f\"ur Mathematische Physik, Mendelssohnstra{\ss}e 3, 38106 Braunschweig, Germany}
\author{C. Kl\"ockner}
\affiliation{Technische Universit\"at Braunschweig, Institut f\"ur Mathematische Physik, Mendelssohnstra{\ss}e 3, 38106 Braunschweig, Germany}

\author{D.M.\ Kennes}
\affiliation{Institut f\"ur Theorie der Statistischen Physik, RWTH Aachen University and JARA-Fundamentals of Future Information Technology, 52056 Aachen, Germany}

\affiliation{Max  Planck  Institute  for  the  Structure  and  Dynamics  of  Matter,Luruper  Chaussee  149,  22761  Hamburg,  Germany}

\author{C.\ Karrasch}
\affiliation{Technische Universit\"at Braunschweig, Institut f\"ur Mathematische Physik, Mendelssohnstra{\ss}e 3, 38106 Braunschweig, Germany}

\setlength{\abovedisplayskip}{4pt}
\setlength{\belowdisplayskip}{4pt}  

\date{\today}

\begin{abstract}
We recapitulate recent developments of the functional renormalization group (FRG) approach to the steady state of systems out of thermal equilibrium. In particular, we discuss second-order truncation schemes which account for the frequency-dependence of the two particle vertex and which incorporate inelastic processes. Our focus is on two different types of one-dimensional fermion chains: i) infinite, open systems which feature a translation symmetry, and ii) finite systems coupled to left and right reservoirs. In addition to giving a detailed and unified review of the technical derivation of the FRG schemes, we briefly summarize some of the key physical results. In particular, we compute the non-equilibrium phase diagram and analyze the fate of the Berezinskii-Kosterlitz-Thouless transition in the infinite, open system.

\end{abstract}

\maketitle

\section{INTRODUCTION} 

Correlations play a vital role in low-dimensional quantum many body systems, and their accurate theoretical description requires sophisticated methods. Treating a system that is not in thermal equilibrium is particularly challenging. For quantum impurity problems (i.e., in zero dimensions), a variety of powerful tools have been developed \cite{Eckel2010}. These include the numerical renormalization group \cite{Anders2005}, tensor network approaches \cite{Schmitteckert2004,Heidrich-Meisner2009}, quantum Monte Carlo \cite{Han2007,Weiss2008,Schiro2009,Werner2009}, Wegner's flow equations \cite{Kehrein2006}, the perturbative renormalization group \cite{PhysRevLett.87.156802,PhysRevLett.97.236808}, the real-time renormalization group \cite{PhysRevLett.84.3686,Schoeller2009,Lindner2019}, and the functional renormalization group \cite{Metzner2012}.

The situation becomes more complicated in one-dimensional non-equilibrium, in particular if one is interested not only in the transient short to intermediate time dynamics but the steady state of an open system. Tensor networks can reliably simulate the time evolution (and in certain cases also the steady state) but are generally limited by the amount of entanglement and thus short time scales \cite{Daley2004,White2004,Vidal2004,Schollwoeck2011}. Exact diagonalization is feasible only for small, closed systems or open systems modelled by Lindblads (which constitutes an additional approximation). Iterative path integral or quantum Monte Carlo based approaches are generically restricted to small or intermediate interactions and short times \cite{Weiss2008,Muehlbacher2008,Schiro2009,Segal2010}. Bethe-ansatz based approaches have been generalized to non-equilibrium but can only address closed, integrable systems \cite{Bertini2016,Castro-Alvaredo2016}. Methods based on Keldysh-Schwinger Green's functions are not limited by dimensionality and can treat open systems but require a reliable way to compute the self-energy \cite{Rammer1986,Sieberer_2016}. One route is provided by perturbation theory \cite{PuigVonFriesen}, but high-order expansions are demanding. Dynamical mean-field theory  yields a local approximation for the self-energy in the thermodynamic limit and for long times \cite{RevDMFT1,Aoki2014}. However, this is strictly controlled only in infinite dimensions, and accounting for non-local terms via cluster expansions is demanding.

The functional renormalization group recasts a given many-body problem analytically in terms of flow equations for correlation functions \cite{Kopietz2010,Metzner2012,Berges2012,Dupuis_2021}. A common approach is to focus on the flow of vertex functions such as the self-energy or the effective two-particle vertex. The method is set up by introducing an artifical, infrared cutoff $\Lambda$ into the non-interacting Green's function. All vertex functions become $\Lambda$-dependent, and their flow equations can be derived, e.g., using generating functionals. The self-energy can then be obtained by solving the flow equations, and without truncation this procedure is exact. The FRG is not restricted by dimensionality, and reservoirs can be incorporated analytically at no extra cost.

In thermal equilibrium, the FRG can be set up either on the Matsubara or the Keldysh axis. In non-equilibrium, one necessarily needs to work in Keldysh space, and FRG approaches for fermionic problems have been developed to study the steady state \cite{Gezzi2007,Jakobs2007,Karrasch2010,Karrasch2010b,Jakobs2010,Laakso2014,Rentrop2014,Khedri2018,Caltapanides2021,Jakobs,Karrasch}, the real-time dynamics \cite{Kloss2011,Kennes2012,Bar_Lev_2017}, and Floquet setups \cite{AK1,AK2}. Other FRG studies for out-of-equilibrium problems can be found in \cite{Kloss2011,Sieberer2013_prl,Sieberer2014,Pawlowski2015,Sieberer_2016}. 

In order to solve the flow equations in practice, they need to be truncated. The simplest approach is to consider only the flow of the self-energy. The resulting approximation is strictly correct to leading order in the interaction $U$ but contains an infinite number of higher-order terms. In equilibrium, such an approach successfully captures Luttinger-liquid power laws to first order in $U$ \cite{Meden2002,Andergassen2004} or transport properties of quantum dots \cite{Karrasch2006,Karrasch2008_2}. In non-equilibrium, one can  tackle power laws in the interacting resonant level model \cite{Karrasch2010,Karrasch2010b,Kennes2012,AK1,AK2,Caltapanides2021} and also address properties of the single impurity Anderson model \cite{Gezzi2007}, Luttinger liquid behavior \cite{Jakobs2007}, or excited states \cite{Klockner_PRB_2018}.

The first-order FRG approximation yields a frequency-independent self-energy which can be interpreted in terms of renormalized system parameters. Inelastic processes are not accounted for. This can be remedied by incorporating the flow of the two-particle scattering (which is the next-higher vertex function); the resulting approach is correct to order $U^2$. In principle, it is straightforward to set up such a scheme. However, the two-particle vertex depends on two ingoing and two outgoing single-particle and frequency indices, which results in a large number of flow equations that are often hard to handle in practice (i.e., in a numerical implementation). Thus, one needs to devise further approximation schemes such as the so-called channel decomposition \cite{Karrasch2008}.

Second-order FRG schemes have been developed for impurity problems both in and out of equilibrium \cite{Hedden2004,Karrasch2008,Jakobs2010,Kinza2013,Laakso2014,Rentrop2014} as well as for one-dimensional systems in equilibrium \cite{Bauer2014,Sbierski2017,Sbierski2018,WeidingerPRB2017,Markhof2018,Weidinger2019}. 
The equilibrium FRG can address second-order Luttinger liquid power laws \cite{Sbierski2017} and captures the phase transition between a Luttinger liquid and a Mott insulating phase \cite{Markhof2018}. 

In two dimensions, equilibrium FRG schemes which account for the flow of the two-particle vertex (but not necessarily the self-energy) are commonly used to study ordering tendencies in Hubbard-type models; see \cite{Zanchi1998,Halboth2000,Honerkamp2001} for early seminal works and \cite{Metzner2012} for a recent review.

In sum, this demonstrates that the FRG is a flexible approach which is not limited by the dimensionality of the system at hand, which can access large systems and large times, is not bound by the amount of entanglement, and which can treat open systems. However, in the context discussed here the method is perturbatively motivated in the strength of the two-body interaction $U$, and the results are strictly controlled only to the order of the truncation (e.g., $U$ or $U^2$). This issue needs to be addressed for each problem under investigation. In a nutshell, the FRG can play out its strengths when one is interested in studying complex setups that cannot be treated using more accurate methods (such as tensor networks).

Second-order FRG schemes which can treat one-dimensional systems out of thermal equilibrium have only been devised recently \cite{Klockner_PRL,Klockner_NJP,Klockner_PRB_2020,Klockner_PhD_thesis}. It is the aim of this paper to review these developments, to present technical details of the different works in a unified fashion, and to briefly recapitulate the main results as well as the failures of the method.

In Ref.~\onlinecite{Klockner_NJP}, a second-order FRG approach was developed to study the steady state of infinite one-dimensional chains which are translation-invariant up to shifts in on-site energy. In addition to devising a meaningful truncation scheme, this also requires dealing with Green's functions of infinite systems. The method was subsequently applied to investigate the non-equilibrium phase diagram of a chain that is coupled to a substrate (i.e., reservoirs) and that is driven out of equilibrium by a longitudinal electrical field (a generalized Wannier-Stark ladder). One should note that such a setup cannot be tackled easily using other approaches such as tensor networks and is thus well-suited for a FRG treatment. Results are published both in Ref.~\onlinecite{Klockner_NJP} and Ref.~\onlinecite{Klockner_PRL}.

In Ref.~\onlinecite{Klockner_PRB_2020}, a second-order FRG framework was developed to study the steady state of finite one-dimensional system which are driven out of equilibrium by (usually left and right) reservoirs. Using parallelization over hundreds of compute nodes, one can access the steady-state of chains of up to $\sim$50 sites. It turns out, however, that this setup behaves non-perturbatively and is not susceptible to a FRG treatment.

This exposition is structured as follows. After specifying the model in Sec.~\ref{sec:Hamiltonian}, we recapitulate the Keldysh formalism and also discuss details on how to deal with infinite systems (Sec.~\ref{sec:gf_keldysh}). We then introduce the general FRG formalism in the common framework (e.g., the cutoff or channel decomposition) that underlies both the infinite and the finite system (Sec.~\ref{sec:FRG}). We move forward by discussing specifics of the FRG implementation for the infinite, translation-invariant case as well as a few exemplary results (Secs.~\ref{sec:app_inf} and \ref{sec:inf_sys_res}). We proceed similary for the finite system (Secs.~\ref{sec:app_finite} and \ref{sec:fin_sys_res}).

The original works that this review is based on can be found in Refs.~\onlinecite{Klockner_NJP, Klockner_PRB_2020, Klockner_PRL, Klockner_PhD_thesis}.

\section{HAMILTONIAN}\label{sec:Hamiltonian}

We will focus on the study of open quantum systems, where the full Hilbert space is a composite space of the system's degrees of freedom and the environment attached to it. The system is allowed to exchange energy and particles with the environment through a coupling term in the Hamiltonian. The total Hamiltonian reads:
\begin{eqnarray}\label{eq:full_H}
H = H_{\text{sys}} + H_{\text{res}}+H_{\text{coup}},~~~
H_{\text{res}} = \sum_{\nu}H_{\text{res}}^{\nu},~~~~
H_{\text{coup}} = \sum_{\nu}H_{\text{coup}}^{\nu}.
\end{eqnarray}
Here $H_{\text{sys}}$ and $H_{\text{res}}^{\nu}$ stand for the independent Hamiltonians of the system and the reservoir labelled by the index $\nu$, respectively, whereas $H_{\text{coup}}^{\nu}$ represents the coupling between the system and a specific reservoir. In what follows, both the system and environment are assumed to contain only fermionic particles following Fermi-Dirac statistics. The microscopic description of the general interacting fermionic system is given in second quantized notation:
\begin{eqnarray}\label{sys_h}
H_{\text{sys}}=\sum_{ i,j}h_{ij}c_{i}^{\dagger}c_{j}+\frac{1}{4}\sum_{ijkl}v_{ijkl}c_{i}^{\dagger}c_{j}^{\dagger}c_{l}c_{k}, \hspace{15pt}
\end{eqnarray}
with the anti-symmetrized interaction tensor $v_{ijkl}=-v_{jikl}=-v_{ijlk}$ and the Hermitian single-particle Hamiltonian $h_{ij}=h_{ji}^{*}$. The single particle indices represent any microscopic degrees of freedom in the system. The operators $c_{i}$ denote annihilation of a fermion with quantum number $i$ and satisfy the anticommutation relations:
\begin{eqnarray}
\left[c_{i},c_{j}^{\dagger}\right]_{+}=c_{i}c_{j}^{\dagger}+c_{j}^{\dagger}c_{i}=\delta_{ij}.
\end{eqnarray}
The environment is represented by a macroscopically large number of degrees of freedom, with each reservoir governed by a non-interacting Hamiltonian of the form:
\begin{eqnarray}\label{res_h}
H_{\text{res}}^{\nu}=\sum_{k} \epsilon^{\nu}_{k}a_{k,\nu}^{\dagger}a_{k,\nu}, 
\end{eqnarray} 
with $k$ representing some microscopic degree of freedom labeling the state. For fermionic baths, one has the anticommutation relations $\big[a_{k,\nu},a_{k'\nu'}^{\dagger}\big]_{+}=\delta_{kk'}\delta_{\nu\nu'}$. Finally, we allow the system and the reservoir $\nu$ to be coupled to each other by a term:
\begin{eqnarray}\label{coup_h}
H_{\text{coup}}^{\nu} = \sum_{k,i}t^{\nu}_{i,k}c_{i}^{\dagger}a_{k,\nu} + \text{H.c.},
\end{eqnarray}
where $t^{\nu}_{i,k}$ are the couplings between the system and the reservoir. 

It will always be assumed that the quantities $h_{ij}$ and $v_{ijkl}$ are finite-ranged. This motivates the introduction of range parameters $R_{h}$ and $R_{v}$ for which
\begin{eqnarray}\label{eq:r_params_h_v}
h_{ij}=0\hspace{5pt}\forall |i-j|\geq R_{h}, \hspace{10pt} v_{ijkl}=0\hspace{5pt}\forall \text{dist}(i,j,k,l)\geq R_{v},
\end{eqnarray}
where $|i-j|$ refers to the distance between single particle indices $i,j$, and $\text{dist}(i,j,k,l)$ represents the pairwise distance between indices.

\section{GREEN FUNCTIONS IN THE KELDYSH FORMALISM}\label{sec:gf_keldysh}
The Keldysh formalism is employed to study the steady state of non-equilibrium quantum systems~\cite{Schwinger_1961, KadanoffBaym_1962,Keldysh1965}. The configuration of the system at some initial time $t_{0}$ corresponds to a non-interacting, equilibrium state. The time evolution is governed by the Hamiltonian $H$, which is time-independent. In order to make this work self-contained, we now recapitulate a few basic concepts.

\subsection{Fundamentals}

\subsubsection{Definitions}

Although different conventions might be found in the literature, here we use the following representation for the Green's functions in the Keldysh basis:
\begin{eqnarray}\label{eq:green_functions_matrix}
G=\left(\begin{matrix}
G^{11} && G^{12}\\
G^{21} && G^{22}
\end{matrix}\right)=\left(\begin{matrix}
G^{\text{ret}} && G^{\K}\\
0 && G^{\text{adv}}
\end{matrix}\right),
\end{eqnarray}
where the indices $\alpha=1,2$ will be termed the \emph{Keldysh indices} in what follows, indicating that the appropiate rotation of the contour operators has already been performed. The fact that $G^{21}=0$ is a consequence of the causality constraint. The components of the retarded propagator are given by:
\begin{eqnarray}\label{gret}
G_{\lambda_1\lambda_2}^{\text{ret}}(t,t')&=&-\mathrm{i}\theta(t-t')\big\langle\big[d_{\lambda_2}^{\dagger}(t'),d_{\lambda_1}(t)\big]_{+}\big\rangle,~~~  d_\lambda\in \{c_i, a_{k,\nu}\},\nonumber\\
G_{\lambda_1\lambda_2}^{\text{ret}}(\omega)&=&\int_{-\infty}^{+\infty}dt e^{i\omega t}G^{\text{ret}}_{\lambda_1\lambda_2}(t)=G^{\text{adv}}_{\lambda_2\lambda_1}(\omega)^{*},
\end{eqnarray}
where the operators $d_\lambda(t)$ are represented in the Heisenberg picture,
and in the last equation the limit $t_{0}\to -\infty$ has been taken so that it is reasonable to assume that $G^{\text{ret}}(t,t')$ solely depends on $t-t'$. The statistical average of operators is defined by:
\begin{eqnarray}\label{eq:tr_av}
\langle ... \rangle= \text{Tr}\left(\hat{\rho}_{0}...\right), \hspace{10pt}\hat{\rho}_{0}=\hat{\rho}_{\text{sys}}\otimes\hat{\rho}_{\text{res}}.
\end{eqnarray}
In this work, we focus on the case where the density matrix $\hat{\rho}_{0}$ of the initial equilibrium configuration does not couple the system and environment parts. Moreover, the initial density matrices $\hat{\rho}_{\text{sys}}$ and $\hat{\rho}_{\text{res}}$ must be each associated with non-interacting Hamiltonians (in order to set up diagrammatics on the Keldysh contour).

The components of the Keldysh propagator are defined by:
\begin{eqnarray}\label{gk}
G^{\K}_{\lambda_1\lambda_2}(t,t') &=& \mathrm{i}\left\langle d_{\lambda_2}^{\dagger}(t')d_{\lambda_1}(t) - d_{\lambda_1}(t)d_{\lambda_2}^{\dagger}(t')\right\rangle,~~~  d_\lambda\in \{c_i, a_{k,\nu}\},\nonumber\\
G^{\K}_{\lambda_1\lambda_2}(\omega) &=& \int_{-\infty}^{+\infty}dt e^{i\omega t}G^{\K}_{\lambda_1\lambda_2}(t).       
\end{eqnarray}

The above definitions for the Green's functions are completely general and hold whether or not interactions are present. In case where interactions in the system are not accounted for, the free Green's functions will be represented by a lowercase letter $g(\omega)$:
\begin{eqnarray}\label{eq:g_comps_keldysh}
g(\omega)=\left(\begin{matrix}
g^{\text{ret}}(\omega) && g^{\K}(\omega)\\
0 && g^{\text{adv}}(\omega)
\end{matrix}\right),\hspace{10pt}v_{ijkl}=0.
\end{eqnarray}
Finally, when referring explicitly to the bare Green functions of a non-interacting, decoupled Hamiltonian, we will add a superscript $g^{0}(\omega)$ to all components of Eq.~\eqref{eq:g_comps_keldysh}:
\begin{eqnarray}\label{eq:g0_comps_keldysh}
g^{0}(\omega)=\left(\begin{matrix}
g^{0,\text{ret}}(\omega) && g^{0,\K}(\omega)\\
0 && g^{0,\text{adv}}(\omega)
\end{matrix}\right),\hspace{10pt} H_{\text{coup}}= 0,\hspace{5pt} v_{ijkl}=0.
\end{eqnarray}
Note that all Green's functions are matrices which carry single-particle indices from the set $\{i, (k,\nu)\}$.

\subsubsection{Bare Green's functions}

The bare Green's functions for the initially isolated system are given by:
\begin{eqnarray}\label{eq:qq_g0_comps}
g^{0,\text{ret}/\text{adv}}_{ij}(\omega)=\frac{1}{\omega -h\pm \mathrm{i}0^{+}}\bigg|_{ij}, \hspace{10pt} g^{0,\text{K}}_{ij}(\omega)=\left(1-2n^\text{s}\right)\left[g^{0,\text{ret}}(\omega)-g^{0,\text{adv}}(\omega)\right]\big|_{ij},
\end{eqnarray}
where $h$ represents the single-particle Hamiltonian matrix with entries $h_{ij}$, and $n^\text{s}$ refers to a matrix containing the statistical occupation numbers in the initial state of the system at $t_{0}$:
\begin{eqnarray}\label{eq:initial_occupations}
n_{ij}^\text{s}=\langle c_{j}^{\dagger}c_{i}\rangle.
\end{eqnarray} 
It is instructive to relate the expressions in Eq.~(\ref{eq:qq_g0_comps}) to the ones in time domain~\cite{Klockner_PRB_2018}:
\begin{eqnarray}
g^{0,\text{ret}}_{ij}(t,t')&=&-\mathrm{i}\theta(t-t')e^{-\mathrm{i}h(t-t')}\big|_{ij},\nonumber\\
g^{0,\K}_{ij}(t,t')&=&-\mathrm{i}e^{-\mathrm{i}h(t-t_{0})}\left(1-2n^\text{s}\right)e^{-\mathrm{i}h(t_{0}-t')}\big|_{ij} \underbrace{=}_{[n^\text{s},h]=0}-\mathrm{i}\left(1-2n^\text{s}\right)e^{-\mathrm{i}h(t-t')}\big|_{ij}.
\end{eqnarray}
We have used that necessarily $[n^\text{s},h]=0$; otherwise $g^{0,\K}(t,t')$ does not become a function of the time difference $t-t'$. Performing a Fourier transform with the appropiate infinitesimal convergence factors $0^{+}$ in the integrals, we obtain:
\begin{eqnarray}\label{eq:fdt}
g^{0,\text{ret}}_{ij}(\omega)&=&\frac{1}{\omega -h + \mathrm{i}0^{+}}\bigg|_{ij},\nonumber\\
g^{0,\K}_{ij}(\omega)&=&\left[1-2n^\text{s}\right]\left(\frac{1}{\omega-h+\mathrm{i}0^{+}}-\frac{1}{\omega-h-\mathrm{i}0^{+}}\right)\bigg|_{ij}\nonumber =  \left[1-2n^\text{s}\right]\left[g^{0,\text{ret}}(\omega)-g^{0,\text{adv}}(\omega)\right]\big|_{ij}.
\end{eqnarray}

For the environment degrees of freedom, the bare Green's functions read:
\begin{eqnarray}\label{eq:res_g0_comps}
g^{0,\text{ret}/\text{adv}}_{(k\nu)(k'\nu')}(\omega)=\frac{1}{\omega -\epsilon^{\nu}_{k}\pm \mathrm{i}0^{+}}\delta_{kk'}\delta_{\nu\nu'}, \hspace{10pt} g^{0,\text{K}}_{(k\nu)(k'\nu')}(\omega)=\left[1-2n^{\nu}(\omega)\right]\left[g^{0,\text{ret}}_{(k\nu)(k'\nu')}(\omega)-g^{0,\text{adv}}_{(k\nu)(k'\nu')}(\omega)\right],
\end{eqnarray}
where we chose the initial statistics to be governed by the Fermi distribution $n^{\nu}(\omega)$:
\begin{eqnarray}\label{eq:fermi_dirac_dist}
n^{\nu}(\omega)=\frac{1}{\text{exp}[(\omega-\mu_{\nu})/T_{\nu}]+1},
\end{eqnarray}
with $\mu_{\nu}$ and $T_{\nu}$ representing the chemical potential and temperature of a specific reservoir $\nu$, respectively.

\subsubsection{Fluctuation-Dissipation Theorem}

In thermal equilibrium, not all three components $G^{\text{ret}},G^{\text{adv}},G^{\K}$ are independent from each other; they are related by the Fluctuation-Dissipation Theorem (FDT)~\cite{AltlandSimons,Kamenev}:
\begin{eqnarray}\label{fdt_theorem}
G^{\K}(\omega) = \left[1-2n(\omega)\right]\left[G^{\text{ret}}(\omega) - G^{\text{adv}}(\omega)\right],
\end{eqnarray} 
where $n(\omega)$ is the Fermi function of a global temperature $T$ and a global chemical potential $\mu$. As one goes away from the equilibrium limit, the FDT is not preserved; however, the relation between different Green's functions can still be investigated by using effective distributions for the non-equilibrium steady state. A common choice is to represent the distribution functions by a Hermitian matrix $1-2n_{\text{eff}}(\omega)$ so that:
\begin{eqnarray}\label{eq:keldysh_sylvester_eq}
G^{\K}(\omega) = G^{\text{ret}}(\omega)\left[1-2n_{\text{eff}}(\omega)\right]-\left[1-2n_{\text{eff}}(\omega)\right]G^{\text{adv}}(\omega),
\end{eqnarray}
with $n_{\text{eff}}(\omega)$ representing an effective distribution of particles in the non-equilibrium, steady state limit.

\subsection{Dyson's equation}\label{subsec:Dyson_eq}
In frequency space, Dyson's equation reads:
\begin{eqnarray}\label{Dyson_eq}
\tilde{G}^{-1}(\omega) = \tilde{g}(\omega)^{-1}-\tilde{\Sigma}(\omega) ~\Leftrightarrow~ \tilde{G}(\omega) = \tilde{g}(\omega) + \tilde{g}(\omega)\tilde{\Sigma}(\omega)\tilde{G}(\omega).
\end{eqnarray}
For the time being, we employ a general notation ($\tilde G$, $\tilde g$, and $\tilde\Sigma$) where the only condition is that the splitting into a `free part' $\tilde g$ and a `self-energy' $\tilde\Sigma$ must be chosen such that Wick's theorem holds. The reason is that we will employ Eq.~(\ref{Dyson_eq}) in two different scenarios, namely (i) to compute the free Green's function $g$ from the bare Green's function $g^0$ in the purely non-interacting case, and (ii) to express the full Green's function $G$ in terms of a self-energy $\Sigma$ associated with the two-body terms.

The self-energy $\tilde{\Sigma}(\omega)$ in Keldysh space takes the form:
\begin{eqnarray}\label{eq:se_keldysh_space}
\tilde{\Sigma}(\omega)=\left(\begin{matrix}
\tilde{\Sigma}^{\text{ret}}(\omega) && \tilde{\Sigma}^{\K}(\omega)\\
0 && \tilde{\Sigma}^{\text{adv}}(\omega)
\end{matrix}\right),
\end{eqnarray}
where $\tilde{\Sigma}^{21}(\omega)=0$ due to the causality constraint. The individual components of Dyson's equation are given by
\begin{eqnarray}\label{eq:G_components}
\tilde{G}^{\text{ret}}(\omega)&=& \frac{1}{\tilde{g}^{\text{ret}}(\omega)^{-1}-\tilde{\Sigma}^{\text{ret}}(\omega)},~~~
\tilde{G}^{\text{adv}}(\omega)= \frac{1}{\tilde{g}^{\text{adv}}(\omega)^{-1}-\tilde{\Sigma}^{\text{adv}}(\omega)},\nonumber\\[1ex]
\tilde{G}^{\K}(\omega) &=& \tilde{g}^{\K}(\omega) + \tilde{g}^{\text{ret}}(\omega)\tilde{\Sigma}^{\text{ret}}(\omega)\tilde{G}^{\K}(\omega) + \tilde{g}^{\text{ret}}(\omega)\tilde{\Sigma}^{\K}(\omega)\tilde{G}^{\text{adv}}(\omega) + \tilde{g}^{\K}(\omega)
\tilde{\Sigma}^{\text{adv}}(\omega)\tilde{G}^{\text{adv}}(\omega)\nonumber\\
&=& \tilde{G}^{\text{ret}}(\omega)\left[\tilde{g}^{\text{ret}}(\omega)^{-1}\tilde{g}^{\K}(\omega)\tilde{g}^{\text{adv}}(\omega)^{-1}+\tilde{\Sigma}^{\K}(\omega)\right]\tilde{G}^{\text{adv}}(\omega).
\end{eqnarray}

\subsection{Integrating out the environment degrees of freedom in the absence of interactions}\label{subsec:proj_technique}
We now employ the Dyson equation to compute $g$ from $g^0$ by setting  $\tilde{G}\to  g$, $\tilde{g}\to g^{0}$, and $\tilde{\Sigma}\to \Sigma^{\text{coup}}$. The self-energy is associated with $H_\text{coup}$. We divide the system and environment single-particle degrees of freedom into two different block sectors $P$ and $Q$, respectively:
\begin{eqnarray} \label{eq:proj_1}
\left(\begin{matrix}
g_{PP} && g_{PQ}\\
g_{QP} && g_{QQ}
\end{matrix}\right)
=\left(\begin{matrix}
g^{0}_{PP} && 0\\
0 && g^{0}_{QQ}
\end{matrix}\right) + \left(\begin{matrix}
g^{0}_{PP} && 0\\
0 && g^{0}_{QQ}
\end{matrix}\right)\left(\begin{matrix}
0 && {\Sigma}_{PQ}^{\text{coup}}\\
{\Sigma}_{QP}^{\text{coup}} && 0
\end{matrix}\right)\left(\begin{matrix}
g_{PP} && g_{PQ}\\
g_{QP} && g_{QQ}
\end{matrix}\right),
\end{eqnarray}
where we avoided writing out the $\omega$-dependence and stress that all quantities still carry a Keldysh index. We have used that $g^{0}_{PQ}=g^{0}_{QP}=0$ since there is no coupling between the system and the environment (it is important to stress that this also holds true for the initial density matrix $\hat{\rho}_{0}$). The fact that $\Sigma^{\text{coup}}$ is associated with $H_{\text{coup}}$ entails $\Sigma_{PP}^{\text{coup}}=\Sigma_{QQ}^{\text{coup}}=0$. Since we are ultimately interested only in the Green's function of the system, we solve for $g_{PP}$:
\begin{eqnarray}\label{eq:proj_2}
g_{PP}^{-1} = (g^{0}_{PP})^{-1} - \Sigma_{PQ}^{\text{coup}}g^{0}_{QQ}\Sigma_{QP}^{\text{coup}}.
\end{eqnarray}
This motivates the introduction of hybridization functions (we now write out the Keldysh structure explicitly):
\begin{eqnarray}\label{eq:res_self_energy}
\left(\begin{matrix}
\Gamma^{\text{ret}} && \Gamma^{\text{K}}\\
0 &&\Gamma^{\text{adv}}
\end{matrix}\right) =\left(\begin{matrix}
\Sigma_{PQ}^{\text{coup},\text{ret}} && 0\\
0 && \Sigma_{PQ}^{\text{coup},\text{adv}}
\end{matrix}\right)\left(\begin{matrix}
g^{0,\text{ret}}_{QQ} && g^{0,\text{K}}_{QQ}\\
0 && g^{0,\text{adv}}_{QQ}
\end{matrix}\right)\left(\begin{matrix}
\Sigma_{QP}^{\text{coup},\text{ret}} && 0\\
0 && \Sigma_{QP}^{\text{coup},\text{adv}}
\end{matrix}\right).
\end{eqnarray}
The retarded and advanced parts of the self-energy are simply given by $\Sigma_{i(k,\nu)}^{\text{coup},\text{ret}}=t_{i,k}^{\nu}$ and $\Sigma_{(k,\nu)i}^{\text{coup},\text{adv}}=t_{i,k}^{\nu*}$. On the other hand, changing the non-interacting Hamiltonian does not yield a Keldysh term \cite{Klockner_PRB_2018}, and thus $\Sigma^{\text{coup},\K}_{PQ}=0$. Using Eq.~(\ref{eq:res_g0_comps}), the different components are explicitly given by:
\begin{eqnarray}\label{hybridizations}
\Gamma^{\text{ret}}_{ij}&=&\sum_{\nu}\Gamma^{\nu,\text{ret}}_{ij}(\omega)=\sum_{k,\nu}t_{i,k}^{\nu}t_{j,k}^{\nu*}\frac{1}{\omega -\epsilon_{k}^{\nu}+\mathrm{i}0^{+}}, \nonumber\\
 \Gamma_{ij}^{\K}(\omega)&=&\sum_{\nu}\Gamma_{ij}^{\nu,\K}(\omega)=\sum_{k,\nu}t_{i,k}^{\nu}t_{j,k}^{\nu*}\left[1-2n^{\nu}(\omega)\right]\left[\frac{1}{\omega -\epsilon_{k}^{\nu}+\mathrm{i}0^{+}}-\frac{1}{\omega -\epsilon_{k}^{\nu}-\mathrm{i}0^{+}}\right].
\end{eqnarray}
We now change the notation; in the following, we use the symbol $g$ with the implicit understanding that it always refers to the Green's function $g_{PP}$ of the system. Since Eq.~(\ref{eq:proj_2}) has again the form of a Dyson equation, we can directly use Eq.~(\ref{eq:G_components}) to read off the individual components:
\begin{equation}\label{non_int_g}
g_{PP}\to g:~~~ g^{\text{ret}}(\omega)=\frac{1}{\omega-h-\Gamma^{\text{ret}}(\omega)},~~~
g^{\K}(\omega) = g^{\text{ret}}(\omega)\Gamma^{\K}(\omega)g^{\text{adv}}(\omega),
\end{equation}
where we have simplified the Keldysh component via Eqs.~\eqref{eq:fdt} and~\eqref{eq:G_components}:
\begin{eqnarray}\label{eq:gk_eq_argument}
g^{\K}(\omega)&=&g^{\text{ret}}(\omega)\left[  g^{0,\text{ret}}(\omega)^{-1}g^{0,\K}(\omega)g^{0,\text{adv}}(\omega)^{-1}+\Gamma^{\K}(\omega)\right]g^{\text{adv}}(\omega)\nonumber\\
&=& g^{\text{ret}}(\omega)\left[  g^{0,\text{ret}}(\omega)^{-1}(1-2n^\text{s})\left[g^{0,\text{ret}}(\omega)- g^{0,\text{adv}}(\omega)\right]g^{0,\text{adv}}(\omega)^{-1}+\Gamma^{\K}(\omega)\right]g^{\text{adv}}(\omega)\nonumber\\
&=&g^{\text{ret}}(\omega)\left[  2\mathrm{i}0^{+}  +\Gamma^{\K}(\omega)\right]g^{\text{adv}}(\omega).
\end{eqnarray}
If both $g^{\text{ret}}(\omega)$ and $g^{\text{adv}}(\omega)$ have finite imaginary parts, taking the limit $0^{+}\to 0$ in the above expression is well defined, and only the second term contributes. The initial occupation $n^\text{s}$ of the system does not enter.

Later on, we will assume that the hybridization functions are finite ranged and thus introduce a corresponding range parameter:
\begin{eqnarray}\label{eq:finite_range_gamma}
\Gamma^{\text{ret}}_{ij}=\Gamma^{\K}_{ij}=0\hspace{10pt}\forall |i-j|\geq R_{\Gamma}.
\end{eqnarray}
In this paper, we will always focus on the special situation of a flat density of states in each reservoir, which is known as the \emph{wide band limit} approximation:
\begin{equation}\label{eq:wide_band_limit}
 \Gamma_{ij}^{\nu,\text{ret}}(\omega) = -\mathrm{i}\Gamma_{ij}^{\nu,\text{ret}},~~~\\
 \Gamma_{ij}^{\nu,\K}(\omega) = -2\mathrm{i}\left[1-2n^{\nu}(\omega)\right]\Gamma_{ij}^{\nu,\text{ret}}.
\end{equation}
For zero or infinite reservoir temperatures (which are the only cases relevant for this work), the Keldysh component takes the form
\begin{equation}\begin{split}
    \label{eq:restrRes}
    \Gamma_{ij}^{\nu,\K}(\omega)&=  -2\mathrm{i}\begin{cases}\text{sgn}(\omega-\mu_\nu)\Gamma_{ij}^{\nu,\text{ret}} & T_\nu=0, \\ 0 & T_\nu=\infty.
    \end{cases}
\end{split}
\end{equation}

\subsection{Incorporating interactions}\label{subsec:full_gf}

The effects of two-body interactions $v_{ijkl}\neq 0$ is expressed in terms of a self-energy via Dyson's equation. As before, we make use of the projection technique, and divide the system and environment degrees of freedom into blocks $P,Q$:
\begin{eqnarray}\label{eq:pq_gf}
G=\left(\begin{matrix}
G_{PP} && G_{PQ}\\
G_{QP} && G_{QQ}
\end{matrix}\right)
=\left(\begin{matrix}
g_{PP} && g_{PQ}\\
g_{QP} && g_{QQ}
\end{matrix}\right) + \left(\begin{matrix}
g_{PP} && g_{PQ}\\
g_{QP} && g_{QQ}
\end{matrix}\right)\left(\begin{matrix}
\Sigma_{PP} && 0\\
0 && 0
\end{matrix}\right)\left(\begin{matrix}
G_{PP} && G_{PQ}\\
G_{QP} && G_{QQ}
\end{matrix}\right).
\end{eqnarray}
The fact that interactions only take place between the system's degrees of freedom entails $\Sigma_{PQ}=\Sigma_{QP}=\Sigma_{QQ}=0$. We are again only interested in the system's Green's function and thus solve for $G_{PP}$:
\begin{eqnarray}\label{eq:G_pp}
G_{PP}=g_{PP} + g_{PP} \Sigma_{PP}G_{PP}.
\end{eqnarray}
The individual components are given by Eq.~(\ref{eq:G_components}):
\begin{eqnarray}\label{Dyson_G_components}
G^{\text{ret}}(\omega) &=& \frac{1}{g^{\text{ret}}(\omega)^{-1} - \Sigma^{\text{ret}}(\omega)}=\frac{1}{\omega-h-\Gamma^{\text{ret}}(\omega)-\Sigma^{\text{ret}
}(\omega)},\nonumber\\[1ex]
G^{\K}(\omega) &=& G^{\text{ret}}(\omega)\left[g^{\text{ret}}(\omega)^{-1}g^{\K}(\omega)g^{\text{adv}}(\omega)^{-1}+\Sigma^{\K}(\omega)\right]G^{\text{adv}}(\omega)
=G^{\text{ret}}(\omega)\left[\Gamma^{\K}(\omega) + \Sigma^{\K}(\omega)\right]G^{\text{adv}}(\omega),
\end{eqnarray}
where the notation is again to be understood such that $G$, $g$, and $\Sigma$ only carry the system's single-particle indices $P$. In the last step in Eq.~\eqref{Dyson_G_components}, we have used Eq.~\eqref{non_int_g}.

\subsection{Multi-index notation}
\label{subsec:multi_index}

We make a comment about the notation that will follow throughout this work. Frequently, the Keldysh indices will be merged with the single particle indices in a multi-index notation $1=(\alpha_{1},i_{1})$, so that a single multi-index includes both the Keldysh index and the single particle index. In general, $i_{1}$ might include more than one species of indices; for instance, $i_{1}=(\sigma_{1},k_{1})$ with $\sigma_{1}$ representing a spin degree of freedom, whereas $k_{1}$ can represent an index for spatial or momentum degrees of freedom. In any case, those indices are absorbed in $i_{1}$.

A Keldysh index $\alpha_{1}$ can either refer to the contour basis where $\alpha_{1}\in\{+,-\}$ or the Keldysh basis where $\alpha_{1}\in\{1,2\}$. If a change of basis from the contour to the Keldysh basis takes place (and vice versa), it will always be explicitly mentioned.

A single component of the full Green's functions matrix Eq.~\eqref{eq:green_functions_matrix} is given in this notation by $G_{1'1}(\omega)$. Whenever the full structure of a quantity is needed (rather than its individual components), we will ommit the multi-indices, i.e., the quantity $G(\omega)$ refers to the full Green's function, whereas $G_{1'1}(\omega)$ will refer to the specific component with multi-indices $(1',1)$. The same applies for higher-order structures that might depend on more than a pair of multi-indices. We will sometimes refrain from writing the explicit $\omega$ dependence in the different $G(\omega)$ components; however, this frequency dependence should always be implicitly present in such cases.

As an exception, and \emph{only} in Sec.~\ref{subsec:vertex_eqns_keldysh}, we will conveniently join frequency indices within a multi-index description. In that case, a single multi-index will be defined by $1=(\alpha_{1},i_{1},\omega_{1})$, including the frequency dependence in $\omega_{1}$. 

Finally, whenever a summation is performed over a single multi-index, it is understood that a sum over all indices forming the multi-index is implied.


\subsection{Green's functions in translationally invariant systems}\label{subsec:iter_gf}
In this section, we discuss how to solve the Dyson equation in an infinite system possesing a discrete translational symmetry up to energy shifts $\sim E$ with a unit cell of size $L$. It is assumed that the self-energy $\Sigma$ associated with the interactions is known (e.g., from a FRG treatment). We follow Ref.~\onlinecite{Klockner_NJP} but note that a similar approach was previously proposed in the context of cluster perturbation theory \cite{Neumayer2015}.

\subsubsection{Systems with a discrete translational symmetry}
\label{subsubsec:inf_sys}

We assume that the single-particle indices are spatial indices $i\in\mathbb{Z}$ and that the following relations for $h$ and $v$ hold (the generalization to the case of multiple single-particle indices is straightforward):
\begin{eqnarray}\label{h_E_field}
h_{(i+L)(j+L)} &=& h_{ij}+LE\delta_{ij}, \nonumber\\
v_{(i+L)(j+L)(k+L)(l+L)} &=& v_{ijkl} \hspace{45pt}\forall\hspace{2pt}i,j,k,l\in\mathbb{Z}.
\end{eqnarray}
Here $E>0$ represents an external shift parameter, and the integer $L$ is referred to as the size of the unit cell. An example of a concrete physical system with such properties is the Wannier-Stark ladder~\cite{Wannier1962,Aoki2014,Davison1997,Neumayer2015} Moreover, we demand that the coupling to the reservoirs $\nu\in\mathbb{Z}$ fulfills:
\begin{eqnarray}\label{hyb_trans_inv}
\Gamma^{\nu+L,\text{ret}}_{(i+L)(j+L)}(\omega) &=& \Gamma^{\nu,\text{ret}}_{ij}(\omega-LE) \hspace{15pt}\forall i,j\in\mathbb{Z}\nonumber\\
n^{\nu+L}(\omega)&=& n^{\nu}(\omega-LE)~\Leftrightarrow~ \mu_{\nu+L}=\mu_{\nu}+LE,~T_\nu=T.
\end{eqnarray}      
Note that the Keldysh components $\Gamma^{\nu,\K}_{ij}(\omega)$ also feature this symmetry by virtue of Eq.~\eqref{hybridizations}.

\begin{figure}[t!]
\includegraphics[width=0.4\linewidth,clip]{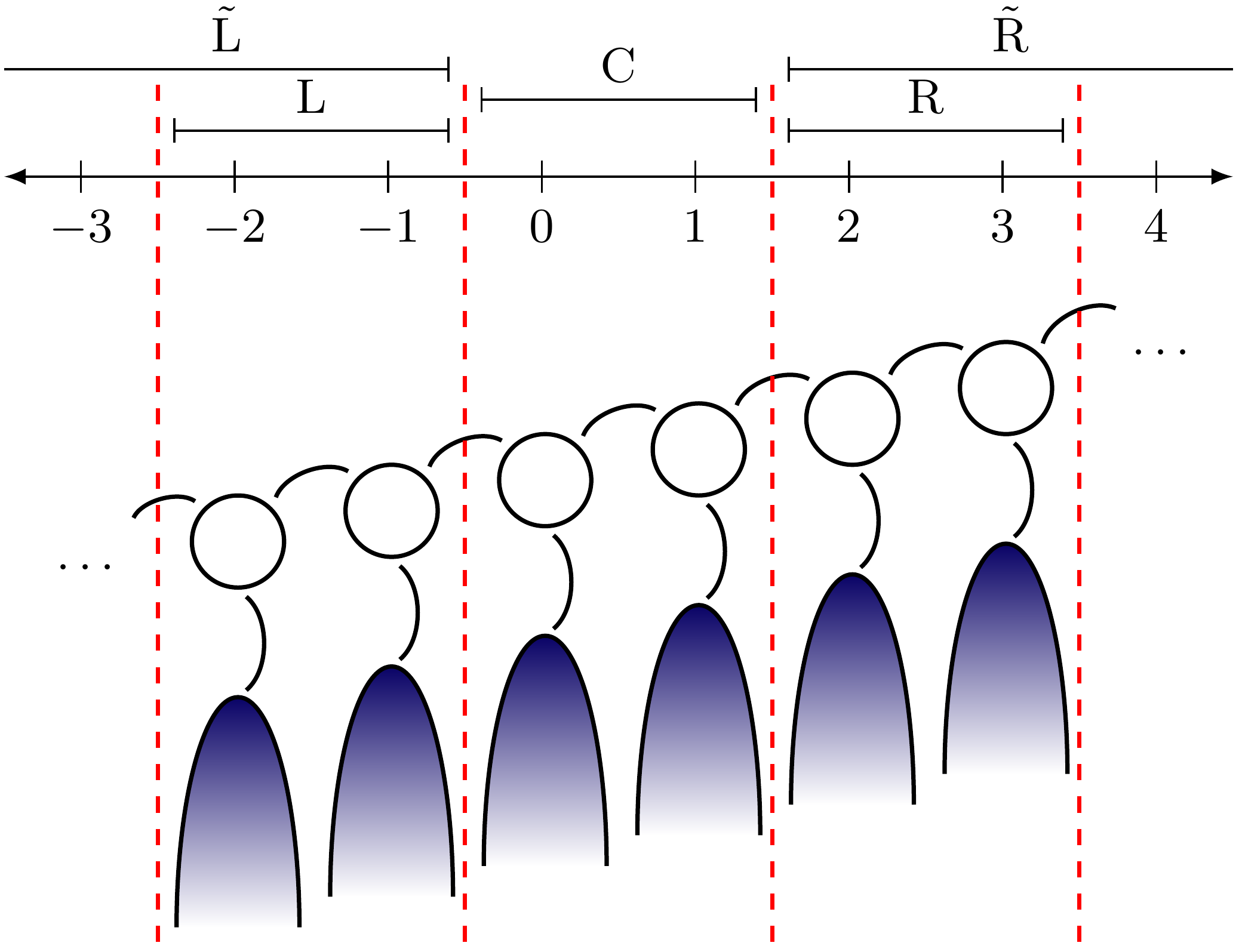}
\caption{(Adapted from Ref.~\onlinecite{Klockner_NJP}, math-mode malformatting fixed.) Illustration of a system with a discrete translational symmetry with $N=2$ (the maximum assumed range of hybridizations and self-energies, see Sec.~\ref{subsubsec:inf_sys}) and a unit cell of $L=1$. In the figure, L, C, and R are regions delimited by the dashed red lines, while $\tilde{\text{L}}, \tilde{\text{R}}$ represent the semi-infinite left and right parts of the chain excluding the central region C [see Eq.~(\ref{eq:regionslcr}) and note that the difference between the region L and the size of the unit cell $L$].}
\label{fig:sys_E}
\end{figure}

The non-interacting Green's functions in Eq.~(\ref{non_int_g}) directly inherit the translational symmetry. The same holds true for the self-energy (which follows from an infinite perturbation series) and thus also for the full Green's function. In a nutshell,
\begin{equation}\begin{split}\label{sym_gsigma}
g_{(1'+L)(1+L)}(\omega) &= g_{1'1}(\omega-LE),\\
\Sigma_{(1'+L)(1+L)}(\omega)&=  \Sigma_{1'1}(\omega-LE),\\
G_{(1'+L)(1+L)}(\omega)&=G_{1'1}(\omega-LE),
\end{split}\end{equation}
where the unit cell shift $(1+L)=(\alpha_{1},i_{1}+L)$ implicitly involves a shift of the single particle index $i_{1}$ without affecting the Keldysh index $\alpha_{1}$. 

While the Dyson equation (\ref{non_int_g}) or (\ref{Dyson_G_components}) cannot be solved in an infinite system, the above symmetries allow us to compute the different components of $G(\omega)$ in a spatially restricted region by applying iterative techniques. Most technical details about how to do this in practice can be found in Ref.~\onlinecite{Klockner_NJP}; however, we also provide them here for ease of reference.

\subsubsection{Further assumptions and notation}

In the rest of this section, it is shown how to solve Dyson's equation (\ref{Dyson_G_components}) for an infinite system under the assumption that the self-energy $\Sigma(\omega)$ is known. Moreover, we demand that
\begin{eqnarray}\label{eq:finite_range_h}
h_{ij}=\Sigma^{\text{ret}/\K}_{ij}(\omega)=\Gamma^{\nu,\text{ret}/\K}_{ij}(\omega)=0 \hspace{10pt} \forall |i-j|\geq N> R_{h},R_{\Gamma},
\end{eqnarray} 
which holds naturally true for $h_{ij}$ and $\Gamma^{\nu,\text{ret}/\K}_{ij}(\omega)$ due to Eqs.~\eqref{eq:r_params_h_v} and~\eqref{eq:finite_range_gamma}, while it constitutes an approximation for the self-energy components $\Sigma^{\text{ret}/\K}_{ij}(\omega)$, which are a priori not of finite range. If our infinite system is characterized by a unit cell of length $L$, we must choose $N\geq L$, with $L$ dividing $N$.

We want to obtain components of the full Green's functions in a region limited by $N$. This motivates us to introduce the following notation for the single-particle indices (see Fig.~\ref{fig:sys_E}):
\begin{equation}\label{eq:regionslcr}\begin{split}
i\in\lli ~~& \Leftrightarrow~~ \hspace*{1.22cm} i \leq -1, \\
i\in\li ~~& \Leftrightarrow~~ -N \leq  i \leq -1, \\
i\in\ci ~~& \Leftrightarrow~~ \hspace*{0.58cm}0 \leq  i \leq N-1, \\
i\in\ri ~~& \Leftrightarrow~~ \hspace*{0.43cm}N \leq  i \leq 2N-1, \\
i\in\rri ~~& \Leftrightarrow~~\hspace*{0.43cm}N \leq  i . \\
\end{split}\end{equation}
Moreover, the notation $G^\textnormal{ret}_{\ci\li}(\omega)$ refers to the retarded Green's function $G^\textnormal{ret}_{(i\in\ci)(j\in\li)}(\omega)$, which is a matrix of size $N\times N$ (and likewise for all other combinations of indices). The indices $\lli,\rri$ are associated with the semi-infinite systems to the left and right of the central region, respectively, whereas $\li,\ri$ refer to the immediate neighboring regions. This implies that any quantity carrying indices $\lli,\rri$ is of infinite dimension, whereas any quantities carrying solely $\ci,\li$, or $\ri$ indices are finite size matrices of dimension $N\times N$. A similar notation is used for any other quantities that depend on two single-particle indices. It is convenient to use a shorthand notation for summation over repeated indices, e.g.,
\begin{equation}\begin{split}
G^\textnormal{ret}_{\ci\li}(\omega) \Sigma^\textnormal{K}_{\li\li}(\omega) =
\sum_{j\in\li} \left[G^\textnormal{ret}_{ij} (\omega)\Sigma^\textnormal{K}_{jk}(\omega)\right]_{i\in\ci, k \in\li}.
\end{split}\end{equation}
Using this notation, the translation symmetry in Eq.~(\ref{sym_gsigma}) takes the form
\begin{equation}\label{eq:iter_translation}
 G^\textnormal{ret/adv/K}_{\li\li}(\omega-NE) = G^\textnormal{ret/adv/K}_{\ci\ci}(\omega) = G^\textnormal{ret/adv/K}_{\ri\ri}(\omega+NE),
\end{equation}
and similarly for the self-energy terms.

\subsubsection{Retarded Green's function}\label{subsubsec:iter_gf1}
We begin with solving for the retarded component of Dyson's equation \eqref{Dyson_G_components} in the central region $\ci$:
\begin{equation}\label{eq:iter_heff}
G^{\text{ret}}(\omega)^{-1} = \omega - h-\Gamma^{\mathrm{ret}}-\Sigma^\mathrm{ret}= T+D,
\end{equation}
where the matrices $T$ and $D$ feature the following block structure:
\begin{equation}\label{block_structure_gret}
    T+D=
  \left(\begin{array}{cc|c|cc}
        &         &     0 &      0 &0 \\
      \multicolumn{2}{c|}{\smash{\raisebox{.5\normalbaselineskip}{$D_{\lli\lli}$}}}
        & T_{\li\ci} &      0 & 0 \\
      \hline \\[-\normalbaselineskip]
      0 &  T_{\ci\li} &     D_{\ci\ci} & T_{\ci\ri} & 0\\
      \hline\\[-\normalbaselineskip]
       0&     0   &     T_{\ri\ci} &       & \\
      
      0 &      0 & 0 & \multicolumn{2}{|c}{\smash{\raisebox{.5\normalbaselineskip}{$D_{\rri\rri}$}}} \\
    \end{array}\right).
\end{equation}
All matrices $D_{\ci\ci}$, $T_{\ci\li}$, $T_{\li\ci}$, $T_{\ci\ri}$, and $T_{\ri\ci}$ are of size $N\times N$, with the matrices forming the $T$ block being in general \emph{not} related by Hermitian conjugation, e.g., \(T_{\li\ci}\neq T_{\ci\li}^\dag\) due to the inclusion of the self-energy. The frequency dependence will often be ommited to improve readability.

To compute the block components in Eq.~\eqref{block_structure_gret}, the general formula for the inverse of a block matrix is employed
\begin{equation}\label{eq:iter_matinv}\begin{split}
\begin{pmatrix} V & W \\ X & Y \end{pmatrix}^{-1}
= \begin{pmatrix} V^{-1} + V^{-1}WY_IXV^{-1} & -V^{-1}WY_I \\ -Y_IXV^{-1} & Y_I \end{pmatrix},~~~~~
Y_I= (Y-XV^{-1}W)^{-1},
\end{split}\end{equation}
which also admits the form
\begin{equation}\label{eq:iter_matinv2}\begin{split}
\begin{pmatrix} V & W \\ X & Y \end{pmatrix}^{-1}
= \begin{pmatrix} V_I & -V_IWY^{-1} \\ -Y^{-1}XV_I & Y^{-1}+Y^{-1}XV_IWY^{-1} \end{pmatrix},~~~~~
V_I= (V-WY^{-1}X)^{-1}.
\end{split}\end{equation}
Successive application of Eqs.~\eqref{eq:iter_matinv} and (\ref{eq:iter_matinv2}) leads to
    \begin{equation}\label{eq:iter_matinv3}
        \begin{split}
            \left[\left(\begin{array}{c|cc} N & U & 0 \\ \hline V & W & X \\ 0 & Y & Z \end{array}\right)^{-1}\right]_{22}&=
        \left[\begin{pmatrix} W-VN^{-1}U & X\\ Y & Z\end{pmatrix}^{-1}\right]_{11}
        = \left(W-VN^{-1}U-XZ^{-1}Y\right)^{-1},
    \end{split}
    \end{equation}
where the decomposition following the first step identifies the blocks as written in the LHS of the equation.

We only want to compute the retarded Green's function $G^\tn{ret}_{ij}$ in a region where $|i-j|<N$, implying that it is sufficient to determine $G^{\tn{ret}}_{\ci\ci}$, that is, to determine the components of the central block region in Eq.~\eqref{block_structure_gret}. Applying Eq.~(\ref{eq:iter_matinv3}) to Eq.~(\ref{eq:iter_heff}) and using that \(T_{\ci(\lli\setminus\li)}=0\), it is easy to show that
\begin{equation}\label{eq:iter_greta}\begin{split}
& G^{\tn{ret}}_{\ci\ci}(\omega) = \frac{1}{D_{\ci\ci}-T_{\ci\li}[D^{-1}]_{\li\li}T_{\li\ci} -T_{\ci\ri}[D^{-1}]_{\ri\ri}T_{\ri\ci} }.
\end{split}\end{equation}
The above equation has two unknown blocks $[D^{-1}]_{\li\li}$ and $[D^{-1}]_{\ri\ri}$, which correspond to the first and last block of the inverse of the two matrices $D_{\lli\lli}$ and $D_{\rri\rri}$. Note that while $[D^{-1}]_{\li\li}$ and $[D^{-1}]_{\ri\ri}$ are matrices of dimension $N\times N$, they are associated with two semi-infinite systems (represented by the $D_{\lli\lli}$ and  $D_{\rri\rri}$ blocks) and need to be determined iteratively; such iterative procedure will be discussed now.

\subsubsection{Auxiliary retarded Green's function}
\label{subsubsec:iter_gaux_ret}
It is convenient to define the auxiliary Green's function
\begin{equation}
\calrLR=D^{-1} 
\end{equation}
as the inverse of the matrix in Eq.~(\ref{eq:iter_heff}) for $T_{\li\ci}=T_{\ci\li}=T_{\ri\ci}=T_{\ci\ri}=0$, which is now block-diagonal. Moreover, one defines auxiliary functions for the case when only one of the semi-infinite systems is decoupled from the central region. When $T_{\li\ci}=T_{\ci\li}=0$, the left block $D_{\lli\lli}$ is decoupled, and the inverse of Eq.~(\ref{eq:iter_heff}) is represented by $\calrL$. If only $T_{\ri\ci}=T_{\ci\ri}=0$, then the block $D_{\rri\rri}$ is decoupled, and the inverse of Eq.~(\ref{eq:iter_heff}) is represented by $\calrR$. The advanced components are defined as $\calaLR=[\calrLR]^\dagger$, $\calaL=[\calrL]^\dagger$, and $\calaR=[\calrR]^\dagger$. In the presence of finite interactions $\Sigma^\tn{ret}\neq0$, these auxiliary Green functions are not physical Green functions of the Hamiltonian; however, they still inherit all of its symmetries such as translation-invariance:
\begin{equation}\label{eq:iter_translation2}\begin{split}
\calrL_{\li\li}(\omega-NE) &= \calrLR_{\li\li}(\omega-NE) = \calrR_{\ci\ci}(\omega),\\
\calrR_{\ri\ri}(\omega+NE) &= \calrLR_{\ri\ri}(\omega+NE) = \calrL_{\ci\ci}(\omega).
\end{split}\end{equation}
The first line can be understood as follows: By looking at the block structure in Eq.~\eqref{block_structure_gret}, it is easy to see that the lower right $N\times N$ block ($\li$) of an isolated $\lli$ system is, up to a shift in energy, identical to the $N\times N$ block ($\ci$) of a system where $\rri$ is removed, due to the properties in Eqs.~\eqref{h_E_field},~\eqref{hyb_trans_inv}, and \eqref{sym_gsigma}. The second line in Eq.~\eqref{eq:iter_translation2} follows similarly. This yields an iterative equation for $\calrLR_{\li\li}$:
\begin{equation}\label{eq:iter_gf2}\begin{split}
\calrLR_{\li\li}(\omega-NE) = \calrR_{\ci\ci}(\omega) = \frac{1}{D_{\ci\ci}(\omega)-T_{\ci\li}(\omega)\calrLR_{\li\li}(\omega)T_{\li\ci}(\omega)},
\end{split}\end{equation}
where Eq.~(\ref{eq:iter_matinv}) is applied to Eq.~(\ref{eq:iter_heff}) with $T_{\ci\ri}=T_{\ri\ci}=0$. The derivation for $\calrLR_{\ri\ri}(\omega)$ is similar and results in replacing indices $\li\to \ri$ on the RHS, and $\calrLR_{\li\li}(\omega-NE)\to\calrLR_{\ri\ri}(\omega+NE)$ on the LHS.

For $E=0$, Eq.~(\ref{eq:iter_gf2}) is local in \(\omega\) and can be solved by a self-consistency loop. At finite $E$, however, the equation couples Green's functions at different frequencies and thus becomes non-local; it can be solved iteratively starting from a boundary condition (details can be found in Ref.~\onlinecite{Klockner_NJP}).

\subsubsection{Keldysh Green's function}\label{subsubsec:iter_gf2}

The Keldysh Green function $G^\tn{K}_{\ci\ci}(\omega)$ is calculated in the central region in the same spirit as in the previous section. The Dyson equation \eqref{Dyson_G_components} for $G^{\K}_{\ci\ci}(\omega)$ reads
\begin{equation}
\begin{split}
& G^\tn{K}_{\ci\ci}(\omega) = \sum_{\si,\si'=\lli,\ci,\rri}G^\tn{ret}_{\ci\si}(\omega)
\Big[\sum_{\nu\in\mathbb{Z}}\Gamma^{\nu,\mathrm{K}}_{\si\si'} +  \Sigma^\mathrm{K}_{\si\si'}\Big] G^\tn{adv}_{\si'\ci}(\omega).
\end{split}
\end{equation}
Using the lower-left component of Eq.~(\ref{eq:iter_matinv}),
\begin{equation}\label{eq:iter_dyson2a}\begin{split}
G^{\tn{ret}}_{\ci\lli} &= - G^{\tn{ret}}_{\ci\ci}T_{\ci\li}[D^{-1}]_{\li\lli}= - G^{\tn{ret}}_{\ci\ci}T_{\ci\li}\calrLR_{\li\lli}, \\
G^{\tn{ret}}_{\ci\rri} &= - G^{\tn{ret}}_{\ci\ci}T_{\ci\ri}[D^{-1}]_{\ri\rri}= - G^{\tn{ret}}_{\ci\ci}T_{\ci\ri}\calrLR_{\ri\rri},
\end{split}\end{equation}
the equation can be written in terms of the auxiliary Green functions:
\begin{equation}\label{eq:iter_gk1}
\begin{split}
 G^\tn{K}_{\ci\ci} = G^\tn{ret}_{\ci\ci}\Big[ \tsk_{\ci\ci} - & T_{\ci\li}\calrLR_{\li\li}\tsk_{\li\ci} - \tsk_{\ci\li} \calaLR_{\li\li} T^\dagger_{\li\ci} \\
 -&T_{\ci\ri}\calrLR_{\ri\ri}\tsk_{\ri\ci}  - \tsk_{\ci\ri} \calaLR_{\ri\ri} T^\dagger_{\ri\ci}
  + T_{\ci\li}\calkLR_{\li\li} T^\dagger_{\li\ci}+ T_{\ci\ri}\calkLR_{\ri\ri}T^\dagger_{\ri\ci}\Big]G^\tn{adv}_{\ci\ci},
\end{split}
\end{equation}
where we have defined
\begin{equation}\label{eq:tsk}
\tsk=\sum_{\nu\in\mathbb{Z}}\Gamma^{\nu,\mathrm{K}} +  \Sigma^\mathrm{K}
\end{equation}
as well as the (Keldysh) auxiliary Green's functions
\begin{equation}\begin{split}
\calkLR_{\li\li}(\omega) = \calrLR_{\li\lli}(\omega)\tsk_{\lli\lli}(\omega)\calaLR_{\lli\li}(\omega),~~~~~
\calkLR_{\ri\ri}(\omega) = \calrLR_{\ri\rri}(\omega)\tsk_{\rri\rri}(\omega)\calaLR_{\rri\ri}(\omega).
\end{split}\end{equation}
In Eq.~(\ref{eq:iter_gk1}), it was used that $\tsk_{\lli\rri}=\tsk_{\ci(\lli\setminus\li)}=\tsk_{\ci(\rri\setminus\ri)}=0$ holds per the assumption in Eq.~\eqref{eq:finite_range_h}. Only $\calkLR_{\li\li}(\omega)$ and $\calkLR_{\ri\ri}(\omega)$ need to be determined, since $G^\tn{ret}_{\ci\ci}$, $\calrLR_{\li\li}$, and $\calrLR_{\ri\ri}$ have already been calculated in the previous section.  The Keldysh auxiliary Green functions $\calkLR_{\li\li}$ and $\calkLR_{\ri\ri}$ can be computed iteratively, exploiting translation-invariance as in the case of the auxiliary retarded Green functions.

\subsubsection{Iterative solution for auxiliary Keldysh Green's function}
\label{subsubsec:iter_gk}

To calculate the auxiliary Green functions $\calkLR_{\li\li}$ and $\calkLR_{\ri\ri}$, the following relation is exploited
\begin{equation}\label{eq:iter_dyson2b}\begin{split}
\calrR_{\ci\lli} = - \calrR_{\ci\ci}T_{\ci\li}\calrLR_{\li\lli},~~~~~~~~
\calrL_{\ci\rri} = - \calrL_{\ci\ci}T_{\ci\ri}\calrLR_{\ri\rri},
\end{split}\end{equation}
which is similar to Eq.~(\ref{eq:iter_dyson2a}) and which follows by using Eq.~(\ref{eq:iter_matinv}) with Eq.~(\ref{eq:iter_heff}) at $T_{\ci\ri}=T_{\ri\ci}=0$. Moreover, translation-invariance is exploited for $\tsk$ [see Eq.~\eqref{sym_gsigma}] and for ${\cal G}^{\tn{ret/adv}}$, leading to:
\begin{equation}\label{eq:iter_translation3}\begin{split}
\calrLR_{\li\lli}(\omega-NE) = \calrR_{\ci(\ci\cup\lli)}(\omega),~~~~~~~~
\calrLR_{\ri\rri}(\omega+NE) = \calrL_{\ci(\ci\cup\rri)}(\omega),
\end{split}\end{equation}
by the same arguments employed to derive Eq.~(\ref{eq:iter_translation2}). The final result for $\calkLR_{\li\li}(\omega-NE)$ reads
\begin{equation}\label{eq:iter_gk2}
\begin{split}
&~ \calkLR_{\li\li}(\omega-NE)\\
=\hspace*{0.1cm} &~  \calrLR_{\li\lli}(\omega-NE)\tsk_{\lli\lli}(\omega-NE)\calaLR_{\lli\li}(\omega-NE)\\
\stackrel{\tn{(\ref{eq:iter_translation3})}}{=}&~ \calrR_{\ci(\ci\cup\lli)}(\omega)\tsk_{(\ci\cup\lli)(\ci\cup\lli)}(\omega)\calaR_{(\ci\cup\lli)\ci}(\omega)\\
\stackrel{\tn{(\ref{eq:iter_dyson2b})}}{=} &~ \calrR_{\ci\ci}\Big[\tsk_{\ci\ci} - T_{\ci\li}\calrLR_{\li\li}\tsk_{\li\ci}  - \tsk_{\ci\li}\calaLR_{\li\li} T^\dagger_{\li\ci} 
+ T_{\ci\li}\underbrace{\calrLR_{\li\lli}\tsk_{\lli\lli}\calaLR_{\lli\li}}_{=\calkLR_{\li\li}(\omega)} T^\dagger_{\li\ci}\Big]\calaR_{\ci\ci}.
\end{split}
\end{equation}
One should not forget that all quantities appearing on the RHS carry a frequency argument $\omega$, which has been omitted to improve readability. A similar equation for $\calkLR_{\ri\ri}$ is obtained in the same way, with the appropiate change of indices and frequency shifts. Eq.~(\ref{eq:iter_gk2}) is of the same form as Eq.~(\ref{eq:iter_gf2}), which means that its solution can be found by a self consistent loop if \(E=0\), or by iterating from the boundary conditions \(\calkLR(\omega)\to 0\) for \(\omega\to \pm\infty\) when \(E\neq 0 \).

\section{FUNCTIONAL RENORMALIZATION GROUP}\label{sec:FRG}
We recapitulate the functional renormalization group (FRG) method \cite{Kopietz2010,Metzner2012} to obtain the self-energy of an interacting system. In particular, we will focus on the FRG on a Keldysh contour formulation \cite{Gezzi2007,Jakobs2007,Karrasch2010,Karrasch2010b,Jakobs2010,Kennes2012,Laakso2014,Rentrop2014,Weidinger2019,Jakobs,Karrasch}.

The key ingredient of the FRG approach is the artificial introduction of a cut-off parameter $\Lambda$ in the non-interacting Green's fucntions of the theory $g^{\Lambda}(\omega)$ which is commonly chosen such that
\begin{equation}\label{eq:cut_off_cond}
\lim_{\Lambda\to\infty}g^{\Lambda}(\omega)=0,~~\lim_{\Lambda\to0}g^\Lambda(\omega)=g(\omega).
\end{equation}
In the Keldysh formalism, this is achieved by introducing an artificial cut-off within the non-interacting Green's functions $g^{\text{ret},\Lambda}(\omega)$, $g^{\text{adv},\Lambda}(\omega)$, and $g^{\K,\Lambda}(\omega)$. Through this dependence on the external cut-off $\Lambda$ (which is also termed flow parameter), the chosen functional becomes scale dependent, too. By studying variations of a given generating functional with respect to $\Lambda$, a set of coupled ordinary differential equations (ODEs) system known as flow equations can be derived. The solution of the flow equations gives direct access to the renormalized components of the functional's expansion terms, which usually represent either the irreducible correlations or the irreducible vertex functions of the interacting system.

The actual derivation of the flow equations depends on the specific form of the generating functional. Here, we will focus on the set of flow equations that provide direct access to the single-particle irreducible $n$-particle vertex functions (a two-particle irreducible scheme can be found in \cite{Dupuis2005,Rentrop2015}; for multi-loop approaches see \cite{Kugler2018,Kugler2018_prl,Tagliavini2019,Yirga2021}). In other words, the chosen generating functional is not a generating functional for the (connected) correlation functions, but rather of the renormalized interaction vertex terms. The flow equations are ODEs of the form (for any $n\in\mathbb{N}$):
\begin{eqnarray}\label{ODE_frg}
\frac{\mathrm{d}\gamma^{\Lambda,(n)}}{\mathrm{d}\Lambda} = \mathcal{F}_{n}\left(\Lambda,\gamma^{\Lambda,(1)},...,\gamma^{\Lambda,(n+1)}\right),
\end{eqnarray}
where $\gamma^{\Lambda,(n)}$ represents the $n$-th particle irreducible vertex, which contains a total of $2n$ single-particle and Keldysh indices, along with $2n$ time or real frequency arguments. For the time being, and following the notation conventions described in Sec.~\ref{subsec:multi_index}, multi-indices also include the frequency arguments. The $n$-particle irreducible vertex function is expressed in terms of $2n$ multi-indices:
\begin{eqnarray}
\gamma^{\Lambda,(n)}= \gamma_{1'2'...n',12...n}^{\Lambda,(n)}\hspace{5pt}\forall n\in\mathbb{N}.
\end{eqnarray}
The flow equations have a hierarchical structure, meaning that the $n$-th component of the vertex is directly coupled to the $n+1$ component by:
\begin{eqnarray}\label{frg_odes}
\frac{\mathrm{d}\gamma^{\Lambda,(1)}}{\mathrm{d}\Lambda} &=& \mathcal{F}_{1}\left(\Lambda,\gamma^{\Lambda,(1)},\gamma^{\Lambda,(2)}\right),\nonumber\\
&\vdots &\nonumber\\
\frac{\mathrm{d}\gamma^{\Lambda,(n)}}{\mathrm{d}\Lambda}&=&\mathcal{F}_{n}\left(\Lambda,\gamma^{\Lambda,(1)},...,\gamma^{\Lambda,(n)},\gamma^{\Lambda,(n+1)}\right),\nonumber\\
\frac{\mathrm{d}\gamma^{\Lambda,(n+1)}}{\mathrm{d}\Lambda}&=&\mathcal{F}_{n+1}\left(\Lambda,\gamma^{(1)},...,\gamma^{\Lambda,(n+1)},\gamma^{\Lambda,(n+2)}\right),\nonumber\\
&\vdots&
\end{eqnarray}       
The most important property of this ODEs system is that it is exact, since the derivation of the flow equations does not rely on any approximations; it is just an exact relation followed by the infinite series expansion in the functional derivatives. The one and two-particle irreducible vertex functions $\gamma^{\Lambda,(1)},\gamma^{\Lambda,(2)}$ are related to the renormalization of the self-energy $\Sigma$ and the bare two-particle vertex $v$, respectively; they describe the flow as one successively includes more low-energy degrees of freedom. For practical applications, the flow equations need to be simplified by some truncation scheme.

\begin{figure}[t!]
    \includegraphics[width=0.8\linewidth,clip]{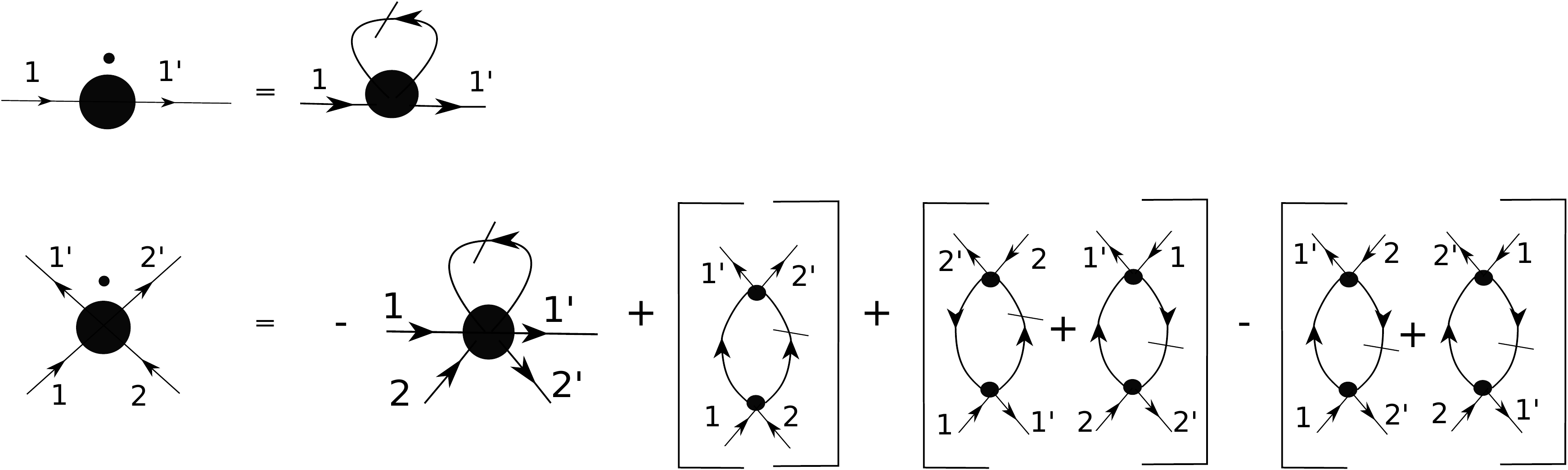}
    \caption{Diagrammatic representation of the flow equations \eqref{eq:flow_eq_contour}. Arrowed lines with a cut represent the single-scale propagator $S^{\Lambda}$ defined in Eq.~\eqref{eq:single_scale_prop}. For the two-particle vertex flow equation, the bracketed terms represent the diagrams associated with to the $\mathrm{p},\mathrm{x},\mathrm{d}$ channels in the channel decomposition (see Sec.~\ref{subsec:2nd_order_approxs}).
    }
    \label{fig:diagrams}
\end{figure}

In the study of the non-equilibrium steady state limit of interacting quantum systems, the flow equations are derived on the Keldysh contour \cite{Gezzi2007,Jakobs2007}. The exact flow equations for the one and two-particle irreducible vertex functions are given by \cite{Metzner2012,Karrasch}:
\begin{eqnarray}\label{eq:flow_eq_contour}
\partial_\Lambda{\gamma}_{1'1}^{\Lambda,(1)}=&-&\sum_{22'}S^{\Lambda}_{22'}\gamma_{1'2'12}^{\Lambda,(2)},\nonumber\\
\partial_\Lambda{\gamma}_{1'2'12}^{\Lambda,(2)}=&-&\sum_{33'}S_{33'}^{\Lambda}\gamma_{1'2'3'123}^{\Lambda,(3)} + \sum_{33'44'}S_{33'}^{\Lambda}G_{4'4}^{\Lambda}\left[\gamma_{1'2'4'3}^{\Lambda,(2)}\gamma_{3'412}^{\Lambda,(2)}\right]\nonumber\\
&+&\sum_{33'44'}S_{33'}^{\Lambda}G_{44'}^{\Lambda}\left[\gamma_{1'3'14}^{\Lambda,(2)}\gamma_{2'4'23}^{\Lambda,(2)} + \gamma_{2'3'24}^{\Lambda,(2)}\gamma_{1'4'13}^{\Lambda,(2)}\right]
-\sum_{33'44'}S_{33'}^{\Lambda}G_{44'}^{\Lambda}\left[\gamma_{2'3'14}^{\Lambda,(2)}\gamma_{1'4'23}^{\Lambda,(2)} + \gamma_{1'3'24}^{\Lambda,(2)}\gamma_{2'4'13}^{\Lambda,(2)}\right],
\end{eqnarray}
which is again visualized in Fig.~\ref{fig:diagrams}. The single-scale propagator $S^{\Lambda}_{22'}$ is defined in terms of the Green's functions as:
\begin{eqnarray}\label{eq:single_scale_prop}
S^{\Lambda}_{22'}=-\sum_{33'}G^{\Lambda}_{23}\left\{\partial_{\Lambda}\left[g^{\Lambda}\right]^{-1}\right\}_{33'}G^{\Lambda}_{3'2'}=\partial_{\Lambda}^{*}G_{22'}^{\Lambda}.
\end{eqnarray} 
In Eq.~\eqref{eq:single_scale_prop}, we have introduced the explicit cut-off derivative operator $\partial_{\Lambda}^{*}$ which acts exclusively over the cut-off in the non-interacting Green's functions $g^{\Lambda}(\omega)$, and not over the self-energies (which also become cut-off dependent due to the flow equations). Note how the irreducible three-particle vertex $\gamma_{1'2'3'123}^{\Lambda,(3)}$ couples to $\gamma_{1'2'12}^{\Lambda,(2)}$ in the RHS of the flow equation for $\gamma_{1'2'12}^{\Lambda,(2)}$; this can be seen from the hierarchical structure of the flow equations in Eq.~\eqref{frg_odes}. Eq.~\eqref{eq:flow_eq_contour} represents the set of flow equations in its most generic form.

\subsection{Vertex flow equations in Keldysh space}
\label{subsec:vertex_eqns_keldysh}

For didactic purposes, we would like to express the flow equation for the two-particle vertex in a more common form; the result can be found in Ref.~\onlinecite{Jakobs2010}. For fermions, any swap in the multi-indices must be associated with a minus sign due to the anticommutativity of the Grassman variables. By changing the multi-index $4\to 4'$ and using $\gamma_{1'2'43}^{\Lambda,(2)}=-\gamma_{1'2'34}^{\Lambda,(2)}$, we obtain
\begin{eqnarray}\label{eq:a_term}
\sum_{33'44'}S_{33'}^{\Lambda}G_{4'4}^{\Lambda}\gamma_{1'2'4'3}^{\Lambda,(2)}\gamma_{3'412}^{\Lambda,(2)} = - \sum_{33'44'}\gamma_{1'2'34}^{\Lambda,(2)}S_{33'}^{\Lambda}G_{44'}^{\Lambda}\gamma_{3'4'12}^{\Lambda,(2)}.
\end{eqnarray}
Likewise, $\gamma_{2'4'23}^{\Lambda,(2)}=\gamma^{\Lambda,(2)}_{4'2'32}$, and together with substituting $3\to 4$ and $3'\to 4'$ in the second term, we obtain
\begin{eqnarray}\label{eq:b_term}
\sum_{33'44'}S_{33'}^{\Lambda}G_{44'}^{\Lambda}\left[\gamma_{1'3'14}^{\Lambda,(2)}\gamma_{2'4'23}^{\Lambda,(2)} + \gamma_{2'3'24}^{\Lambda,(2)}\gamma_{1'4'13}^{\Lambda,(2)}\right]=\sum_{33'44'}\gamma_{1'3'14}^{\Lambda,(2)}\left[S_{33'}^{\Lambda}G_{44'}^{\Lambda}+G_{33'}^{\Lambda}S_{44'}^{\Lambda}\right]\gamma_{4'2'32}^{\Lambda,(2)},
\end{eqnarray}
and similarly:
\begin{eqnarray}\label{eq:c_term}
-\sum_{33'44'}S_{33'}^{\Lambda}G_{44'}^{\Lambda}\left[\gamma_{2'3'14}^{\Lambda,(2)}\gamma_{1'4'23}^{\Lambda,(2)} + \gamma_{1'3'24}^{\Lambda,(2)}\gamma_{2'4'13}^{\Lambda,(2)}\right]=-\sum_{33'44'}\gamma_{1'4'32}^{\Lambda,(2)}\left[S_{33'}^{\Lambda}G_{44'}^{\Lambda}+G_{33'}^{\Lambda}S_{44'}^{\Lambda}\right]\gamma_{3'2'14}^{\Lambda,(2)}.
\end{eqnarray}
A final and important step is to realize that $\gamma^{\Lambda,(2)}$ is closely linked to the two-particle interaction, barring a prefactor of an imaginary unit; this is due to the unitary evolution carried along the Keldysh contour. In particular, the quantum Keldysh action is directly related to the two-particle vertex $\gamma^{\Lambda\to\infty,(2)}_{1'2'12}$ in this limit, since at $\Lambda\to\infty$ the non-interacting propagators $g^{\Lambda\to\infty}(\omega)\to 0$; the only surviving term in the action in such limit is the one proportional to the two-particle interaction terms in Eq.~\eqref{sys_h}. From the derivation of the flow equations, the initial condition for the two-particle vertex reads~\cite{Gezzi2007,Jakobs2007,Jakobs2010}:
\begin{eqnarray}\label{keldysh_bare_vertex}
\gamma_{1'2'12}^{\Lambda\to\infty,(2)}=-\mathrm{i}\begin{cases}
\text{Contour}: \hspace{10pt}v_{1'2'12} =\delta(\omega_{1'}+\omega_{2'}-\omega_{1}-\omega_{2})v_{i_{1'}i_{2'}i_{1}i_{2}}(-\alpha_{1})\delta_{\alpha_{1'}\alpha_{2'}}\delta_{\alpha_{2'}\alpha_{1}}\delta_{\alpha_{1}\alpha_{2}},\\
\text{Keldysh}:\hspace{10pt} \bar{v}_{1'2'12}=\delta(\omega_{1'}+\omega_{2'}-\omega_{1}-\omega_{2})v_{i_{1'}i_{2'}i_{1}i_{2}}\frac{1 - (-1)^{(\alpha_{1'}+\alpha_{2'}+\alpha_{1}+\alpha_{2})}}{4}.
\end{cases}
\end{eqnarray}
In order to avoid problems carrying the imaginary unit, we multiply every term of the two-particle vertex by the imaginary unit making the replacement $\gamma^{\Lambda,(2)}_{1'2'12}\to\mathrm{i}\gamma^{\Lambda,(2)}_{1'2'12}$, so that the initial condition in Eq.~\eqref{keldysh_bare_vertex} now reads:
\begin{eqnarray}\label{eq:gamma2_init_cond}
\lim_{\Lambda\to \infty}\gamma_{1'2'12}^{\Lambda,(2)}=v_{1'2'12}\lor \bar{v}_{1'2'12},
\end{eqnarray}
depending on the chosen basis for the Keldysh indices; in what follows, we will always work in the Keldysh basis. The exact flow equations for the self-energy and the two-particle vertex now read \cite{Jakobs2010}:
\begin{eqnarray}\label{eq:flow_eq_transformed}
\partial_\Lambda{\gamma}^{\Lambda,(1)}_{1'1}&=&\mathrm{i}\sum_{22'}S^{\Lambda}_{22'}\gamma_{1'2'12}^{\Lambda,(2)},\nonumber\\
\partial_\Lambda{\gamma}_{1'2'12}^{\Lambda,(2)}&=&-\mathrm{i}\sum_{33'}S_{33'}^{\Lambda}\gamma_{1'2'3'123}^{\Lambda,(3)}\nonumber
+\mathrm{i}\sum_{33'44'}\gamma_{1'2'34}^{\Lambda,(2)}\left[S_{33'}^{\Lambda}G_{44'}^{\Lambda}\right]\gamma_{3'4'12}^{\Lambda,(2)}\nonumber\\
&&+\mathrm{i}\sum_{33'44'}\gamma_{1'4'32}^{\Lambda,(2)}\left[S_{33'}^{\Lambda}G_{44'}^{\Lambda}+G_{33'}^{\Lambda}S_{44'}^{\Lambda}\right]\gamma_{3'2'14}^{\Lambda,(2)}
-\mathrm{i}\sum_{33'44'}\gamma_{1'3'14}^{\Lambda,(2)}\left[S_{33'}^{\Lambda}G_{44'}^{\Lambda}+G_{33'}^{\Lambda}S_{44'}^{\Lambda}\right]\gamma_{4'2'32}^{\Lambda,(2)}.
\end{eqnarray}

\subsection{Truncation schemes}\label{subsec:trunc_schemes}

Since the system of coupled ODEs in Eq.~\eqref{frg_odes} is infinite, a truncation procedure needs to be devised. Here we briefly describe the two most commonly employed truncation schemes so far encountered in the literature. We restore now the original notation by writing the explicit frequency dependence in all quantities; the vertex functions components read:
\begin{eqnarray}
\gamma_{1'1}^{\Lambda,(1)}\to \gamma_{1'1}^{\Lambda,(1)}(\omega_{1'},\omega_{1});\hspace{5pt}\gamma_{1'2'12}^{\Lambda,(2)}\to\gamma_{1'2'12}^{\Lambda,(2)}(\omega_{1'},\omega_{2'},\omega_{1},\omega_{2}).
\end{eqnarray}

In a truncation scheme of order $m_{c}$, all vertex components with $n>m_{c}$ are set equal to their initial value at the beginning of the flow (for other approaches, see \cite{Weyrauch2008}):
\begin{eqnarray}\label{eq:zero_vertex_conds}
\gamma^{\Lambda,(n>m_{c})}=\lim_{\Lambda\to\infty} \gamma^{\Lambda,(n>m_{c})}\hspace{10pt}\forall\hspace{5pt}\Lambda.
\end{eqnarray}
In the limit $\Lambda\to\infty$, the only non-zero vertex components correspond to the two-particle vertex $\gamma^{\Lambda,(2)}$ [see Eq.~\eqref{keldysh_bare_vertex}], while the other vertex components of arbitrary order vanish in this limit. In what follows, any discussed truncation scheme is based on Eq.~\eqref{eq:zero_vertex_conds}.

It is important to note that a truncation scheme of order $m_{c}$ is not equivalent to a perturbative expansion up to that order. The later is inherently contained in the FRG flow equations, which are exact up to order $m_{c}$ in the interaction vertex due to $\gamma^{\Lambda,(n)}\sim\bar v^{n}$.

\subsubsection{First order truncation}\label{paragraph:first_order_trunc}
The simplest truncation scheme $m_{c}=1$ involves setting $\gamma^{\Lambda,(n>2)}=0$ and $\gamma^{\Lambda,(n=2)}=\bar{v}$, that is, the two-particle irreducible vertex is set equal to its initial bare value, and therefore does not flow with $\Lambda$; similarly, higher order vertex terms are set to zero at any scale of the flow. In a frequency representation, the first order truncation flow equations with $m_{c}=1$ read:
\begin{eqnarray}\label{eq:frg_trunc_1}
-\partial_{\Lambda}\gamma_{1'1}^{\Lambda,(1)}=\partial_{\Lambda}\Sigma_{1'1}^{\Lambda}=-\frac{\mathrm{i}}{2\pi}\sum_{2,2'}\bar{v}_{1'2'12}\int d\Omega S_{22'}^{\Lambda}(\Omega),
\end{eqnarray} 
where $S^{\Lambda}_{22'}(\Omega)$ is the single-scale propagator as defined in Eq.~\eqref{eq:single_scale_prop}, and $\bar{v}_{1'2'12}$ is given by Eq.~\eqref{keldysh_bare_vertex}. We have further identified the self-energy $\Sigma_{1'1}^{\Lambda}=-\gamma_{1'1}^{\Lambda,(1)}$. The self-energy $\Sigma_{1'1}^{\Lambda}$ enters into its own flow equation through the \text{RHS} of Eq.~\eqref{eq:frg_trunc_1}. Such a first order truncation scheme allows for a straightforward implementation of the flow equations \cite{Gezzi2007,Jakobs2007,Karrasch2010,Karrasch2010b,Kennes2012,AK1,AK2,Rentrop2014,Klockner_PRB_2018,Caltapanides2021}; however, the self-energy remains frequency-independent in this scheme, limiting its range of applicability. On the other hand, a second order truncation scheme, while technically more involved, is important in out of equilibrium problems to describe heating effects.

\subsubsection{Second order truncation}\label{subsubsec:second_order_trunc}

We now set $m_{c}=2$ and study the flow of the self-energy and the two-particle vertex. Frequency conservation is inherited through the flow equations, which can be seen by direct inspection of Eq.~\eqref{eq:flow_eq_transformed}. We thus use the notation:
\begin{eqnarray}\label{eq:w_dep_vertex}\label{eq:2p_vertex_parametrization}
\gamma^{\Lambda,(1)}_{1'1}( \omega_{1'},\omega_{1})&=&-\delta(\omega_{1'}-\omega_{1})\Sigma_{1'1}^{\Lambda}(\omega_{1}),  \nonumber\\
\gamma^{\Lambda,(2)}_{1'2'12}(\omega_{1'},\omega_{2'},\omega_{1},\omega_{2})&=&\delta(\omega_{1'}+\omega_{2'}-\omega_{1}-\omega_{2}) \gamma_{1'2'12}^{\Lambda}(\Pi,X,\Delta),
\end{eqnarray}
where we parametrized the frequency dependence of the two-particle vertex $\gamma_{1'2'12}^{\Lambda}$ by a set of three independent bosonic frequencies $\Pi,X,\Delta$ \cite{Hedden2004,Karrasch2008,Jakobs2010,Sbierski2017,Markhof2018,Bauer2014,Klockner_NJP}:
\begin{equation}
    \label{eq:freq_trafo}
    \begin{split}
        \Pi=\omega_1+\omega_2=\omega_{1'}+\omega_{2'},~~~~~~
        X =\omega_{2'}-\omega_1=\omega_{2}-\omega_{1'},~~~~~~
        \Delta =\omega_{1'}-\omega_1=\omega_{2}-\omega_{2'}.
    \end{split}
\end{equation}
Under this parametrization, the self-energy flow equation in Eq.~\eqref{eq:flow_eq_transformed} takes the form:
\begin{equation}
\label{eq:se_flow_eq_2nd_trunc}
\partial_\Lambda\Sigma^{\Lambda}_{1'1}(\omega)=-\frac{\mathrm{i}}{2\pi} \int d\Omega\sum_{22'} \gamma^{\Lambda}_{1'2'12}(\omega+\Omega, \Omega-\omega, 0) S^\Lambda_{22'}(\Omega).
\end{equation}
For the two-particle vertex, we focus on the first non-vanishing term in Eq.~\eqref{eq:flow_eq_transformed} (note that in this truncation scheme $\gamma_{1'2'3'123}^{\Lambda,(3)}=0$), and frequency conservation yields:
\begin{eqnarray}
\omega_{1'}+\omega_{2'}=\omega_{3}+\omega_{4}, \hspace{10pt} \omega_{3'}+\omega_{4'}=\omega_{1}+\omega_{2}.
\end{eqnarray}
Since $S_{33'}^{\Lambda}$ and $G_{44'}^{\Lambda}$ only have weight if $\omega_{3}=\omega_{3'}$ and $\omega_{4}=\omega_{4'}$, it follows that
\begin{eqnarray}
\omega_{1'}+\omega_{2'}=\omega_{1}+\omega_{2},
\end{eqnarray}  
which is one example for how frequency conservation is preserved by the flow equations. We set now:
\begin{eqnarray}
\omega_{3}=\frac{\Pi}{2}-\Omega,~~~\Rightarrow~~~\omega_{4}=\omega_{1'}+\omega_{2'}-\omega_{3}=\frac{\Pi}{2}+\Omega,
\end{eqnarray}
with $\Omega$ defining an internal frequency integration variable. Following the above parametrization:
\begin{eqnarray}
\gamma_{1'2'34}^{\Lambda}(\omega_{1'}+\omega_{2'},\omega_{2'}-\omega_{3},\omega_{1'}-\omega_{3} ) &=& \gamma_{1'2'34}^{\Lambda}\left(\Pi,\Omega+\frac{X-\Delta}{2},\Omega - \frac{X-\Delta}{2} \right),\nonumber\\
\gamma_{3'4'12}^{\Lambda}(\omega_{3'}+\omega_{4'},\omega_{4'}-\omega_{1},\omega_{3'}-\omega_{1} ) &=& \gamma_{3'4'12}^{\Lambda}\left(\Pi,\frac{X+\Delta}{2}+\Omega,\frac{X+\Delta}{2}-\Omega \right).
\end{eqnarray}
A similar analysis can be carried out for the other terms appearing on the RHS of Eq.~\eqref{eq:flow_eq_transformed}, which leads to \cite{Klockner_NJP}
\begin{widetext}
\begin{equation}
    \label{eq:sord}
    \begin{split}
        \partial_\Lambda \gamma^\Lambda_{1'2'12}(\Pi,X,\Delta)=&\frac{\mathrm{i}}{2\pi}\int d\Omega\sum_{33'44'}\\[2ex]
        &\hspace{-2cm}\Bigg\{\gamma^\Lambda_{1'2'34}\left(\Pi, \Omega+\frac{X-\Delta}{2}, \Omega-\frac{X-\Delta}{2}\right)S^\Lambda_{33'}\left(\frac{\Pi}{2}-\Omega\right)G^\Lambda_{44'}\left(\frac{\Pi}{2}+\Omega\right)\gamma^\Lambda_{3'4'12}\left(\Pi,\frac{X+\Delta}{2}+\Omega, \frac{X+\Delta}{2}-\Omega\right)\\[2ex]
       &\hspace{-2.4cm}+\gamma^\Lambda_{1'4'32}\left(\frac{\Pi+\Delta}{2}+\Omega, X, \frac{\Pi+\Delta}{2}-\Omega\right)\biggl[S^\Lambda_{33'}\left(\Omega-\frac{X}{2}\right)G^\Lambda_{44'}\left(\Omega+\frac{X}{2}\right)+\\[2ex]
        &\hspace{3.86cm}G^\Lambda_{33'}\left(\Omega-\frac{X}{2}\right)S^\Lambda_{44'}\left(\Omega+\frac{X}{2}\right)\biggr]\gamma^\Lambda_{3'2'14}\left(\Omega+\frac{\Pi-\Delta}{2}, X, \Omega- \frac{\Pi-\Delta}{2}\right)\\[2ex]
        &\hspace{-2.4cm}-\gamma^\Lambda_{1'3'14}\left(\Omega+\frac{\Pi-X}{2}, \Omega-\frac{\Pi-X}{2}, \Delta\right)\biggl[S^\Lambda_{33'}\left(\Omega-\frac{\Delta}{2}\right)G^\Lambda_{44'}\left(\Omega+\frac{\Delta}{2}\right)+\\[2ex]
        &\hspace{3.91cm}G^\Lambda_{33'}\left(\Omega-\frac{\Delta}{2}\right)S^\Lambda_{44'}\left(\Omega+\frac{\Delta}{2}\right)\biggr]\gamma^\Lambda_{4'2'32}\left(\frac{\Pi+X}{2}+\Omega, \frac{\Pi+X}{2}-\Omega,\Delta\right)\Bigg\}\\[2ex]
        &\hspace{3.91cm} + \mathcal{O}(\bar{v}^{3}).
    \end{split}
\end{equation}
\end{widetext}
Solving these equations in full generality is numerically demanding. For a $L$-level quantum system, the two-particle vertex carries a total of $(2L)^{4}=16L^{4}$ multi-index components. If one discretizes the frequency space in a numerical solution using $N_{\Omega}$ grid points for each of the bosonic frequencies, one ends up with a total of $16L^{4}N_{\Omega}^{3}$ coupled flow equations, which can in practice only be tackled for impurity problems \cite{Jakobs2010,Rentrop2014,Laakso2014}.

\subsection{Choice of a cut-off and single-scale propagator}\label{subsec:cut_off_choice}\label{subsub:initial_conditions}

Up to now, we have not specified the form of the (artificially introduced) cut-off parameter $\Lambda$, which directly enters in the non-interacting propagators $g^{\Lambda}(\omega)$. Physical results should be unaffected by a specific choice of the cut-off. A natural choice of a cut-off is to couple each site of the system to an artificial wide-band reservoir and to use the coupling as the flow parameter $\Lambda$. This is known in the FRG literature as the hybridization cut-off scheme \cite{Jakobs2010}. The cutoff can be added to the full hybridization of the system given in Eq.~\eqref{hybridizations}, which now becomes
\begin{eqnarray}\label{eq:cut_off_hyb}
\Gamma_{ij}^{\text{ret},\Lambda}(\omega) = \Gamma^{\text{ret}}_{ij}(\omega) -\mathrm{i}\delta_{ij}\Lambda,~~~
\Gamma^{\K,\Lambda}_{ij}(\omega) =  \Gamma^{\K}_{ij}(\omega)-2\mathrm{i}\left[1-2n_{i}^{\text{cut}}(\omega)\right]\delta_{ij}\Lambda,
\end{eqnarray}
where we have used the wide-band limit result of Eq.~(\ref{eq:wide_band_limit}). The particle distribution of the artifical reservoirs is governed by Fermi functions $n^{\text{cut}}_{i}(\omega)$, whose temperature and chemical potential we leave unspecified at this point. An advantage of this cut-off scheme is that it automatically preserves causality at any scale $\Lambda$, and the FDT is satisfied in equilibrium \cite{Jakobs2010c}. The initial conditions read:
\begin{eqnarray}\label{eq:initial_conds}
\lim_{\Lambda\to \infty}\Sigma^{\text{ret},\Lambda}_{i_{1'}i_{1}} = \frac{1}{2}\sum_{k}v_{i_{1'}ki_{1}k},~~~~
\lim_{\Lambda\to \infty}\Sigma^{\text{K},\Lambda}_{i_{1'}i_{1}} = 0,~~~~
\lim_{\Lambda\to \infty}\gamma^{\Lambda}_{1'2'12} = \bar{v}_{1'2'12}.
\end{eqnarray}
The initial bare propagator $\bar{v}_{1'2'12}$ in Keldysh space (after the Keldysh rotation) is given by Eq.~\eqref{keldysh_bare_vertex}.

In order to compute the single-scale propagator
\begin{eqnarray}\label{eq:S_prop_matrix}
\left(\begin{matrix}
S^{\text{ret},\Lambda} && S^{\K,\Lambda}\\
0 && S^{\text{adv},\Lambda}
\end{matrix}\right)=-\left(
\begin{matrix}
G^{\text{ret},\Lambda} && G^{\K,\Lambda}\\
0 && G^{\text{adv},\Lambda}
\end{matrix}
\right)
\left(
\begin{matrix}
\partial_{\Lambda}\left[g^{\text{ret},\Lambda}\right]^{-1} && \partial_{\Lambda}\left[-\Gamma^{\K,\Lambda}\right]\\
0 && \partial_{\Lambda}\left[g^{\text{adv},\Lambda}\right]^{-1}
\end{matrix}
\right)
\left(
\begin{matrix}
G^{\text{ret},\Lambda} && G^{\K,\Lambda}\\
0 && G^{\text{adv},\Lambda}
\end{matrix}
\right),
\end{eqnarray}
we first combine $g^{\text{ret},\Lambda}(\omega)$ in Eq.~\eqref{non_int_g} with Eq.~\eqref{eq:cut_off_hyb}:
\begin{eqnarray}\label{eq:S_prop_ret}
S^{\text{ret},\Lambda}(\omega)=-G^{\text{ret},\Lambda}(\omega)\partial_{\Lambda}\left[g^{\text{ret},\Lambda}(\omega)\right]^{-1}G^{\text{ret},\Lambda}(\omega) = -\mathrm{i}G^{\text{ret},\Lambda}(\omega)G^{\text{ret},\Lambda}(\omega).
\end{eqnarray}
For the Keldysh component, we make use of the form for $G^{\K,\Lambda}(\omega)$ given in Eq.~\eqref{Dyson_G_components}:
\begin{eqnarray}\label{eq:S_prop_K}
S^{\K,\Lambda}&=& -G^{\text{ret},\Lambda}\partial_{\Lambda}\left[g^{\text{ret},\Lambda}\right]^{-1}G^{\K,\Lambda} - G^{\K,\Lambda}\partial_{\Lambda}\left[g^{\text{adv},\Lambda}\right]^{-1}G^{\text{adv},\Lambda}+G^{\text{ret},\Lambda}\left[\partial_{\Lambda}\Gamma^{\K,\Lambda}\right]G^{\text{adv},\Lambda}\nonumber \\
&=&S^{\text{ret},\Lambda}\left[\Gamma^{\K,\Lambda}+\Sigma^{\K,\Lambda}\right]G^{\text{adv},\Lambda} + G^{\text{ret},\Lambda}\left[\Gamma^{\K,\Lambda}+\Sigma^{\K,\Lambda}\right]S^{\text{adv},\Lambda} + G^{\text{ret},\Lambda}\left[\partial_{\Lambda}\Gamma^{\K,\Lambda}\right]G^{\text{adv},\Lambda}\nonumber\\
&=& S^{\text{ret},\Lambda}\left[\Gamma^{\K,\Lambda}+\Sigma^{\K,\Lambda}\right]G^{\text{adv},\Lambda} + G^{\text{ret},\Lambda}\left[\Gamma^{\K,\Lambda}+\Sigma^{\K,\Lambda}\right]S^{\text{adv},\Lambda} -2\mathrm{i}\left[1-2n^{\text{cut}}\right] G^{\text{ret},\Lambda}G^{\text{adv},\Lambda},
\end{eqnarray}
where the matrix $n^{\text{cut}}$ has entries given by:
\begin{eqnarray}\label{eq:ncut_eq}
n^{\text{cut}}_{ij}(\omega) = n_{i}^{\text{cut}}(\omega)\delta_{ij}.
\end{eqnarray}

\subsection{General approximations for the second order flow equations: channel decomposition}\label{subsec:2nd_order_approxs}
We have argued that a full solution of the $m_{c}=2$ truncated flow equations given in Eqs.~\eqref{eq:se_flow_eq_2nd_trunc} and~\eqref{eq:sord} is impractical in most applications, and further approximations need to be devised. A prototypical approach is the so-called channel decomposition \cite{Hedden2004,Karrasch2008,Jakobs2010,Sbierski2017,Markhof2018,Bauer2014,Klockner_NJP}, which we will now recapitulate (for extensions of this scheme, see, e.g., Refs.~\onlinecite{Wentzell2020,Tagliavini2019}):
\begin{equation}\label{eq:channel_decomp_pxd}
\begin{split}
\gamma_{1'2'12}^{\Lambda}(\Pi,X,\Delta)\approx \bar{v}_{1'2'12} + \gamma_{1'2'12}^{\mathrm{p},\Lambda}(\Pi)+\gamma_{1'2'12}^{\mathrm{x},\Lambda}(X)+\gamma_{1'2'12}^{\mathrm{d},\Lambda}(\Delta).
\end{split}
\end{equation}
Note that this is strictly fullfilled within perturbation theory (which we will see later explicitly) and thus correct up to second order in the bare interaction. It will prove advantageous to introduce the notation
\begin{eqnarray}\label{eq:bar_channels}
\bar{\gamma}_{1'2'12}^{\{\mathrm{p},\mathrm{x},\mathrm{d}\},\Lambda}=\bar{v}_{1'2'12} + \gamma^{\{\mathrm{p},\mathrm{x},\mathrm{d}\},\Lambda}_{1'2'12}.
\end{eqnarray}
The self-energy flow equation in Eq.~\eqref{eq:se_flow_eq_2nd_trunc} then takes the form
\begin{equation}\label{eq:ford2}
\begin{split}
    \partial_\Lambda\Sigma^\Lambda_{1'1}(\omega) =-\frac{\mathrm{i}}{2\pi} \int d\Omega\sum_{22'}  S^\Lambda_{22'}(\Omega)
    \Big[\bar v_{1'2'12}+ \gamma^{\mathrm{p},\Lambda}_{1'2'12}(\Omega+\omega)+\gamma^{\mathrm{x},\Lambda}_{1'2'12}(\Omega-\omega)+\gamma^{\mathrm{d},\Lambda}_{1'2'12}(0)\Big],
\end{split}\end{equation}
which one can split into three independent contributions:
\begin{eqnarray}\label{eq:sigma_channel_dec}
\partial_{\Lambda}\Sigma_{1'1}^{\mathrm{p},\Lambda}(\omega)&=&-\frac{\mathrm{i}}{2\pi}\sum_{22'}\int d\Omega S_{22'}^{\Lambda}(\Omega)\gamma_{1'2'12}^{\mathrm{p},\Lambda}(\Omega+\omega),\nonumber\\
\partial_{\Lambda}\Sigma_{1'1}^{\mathrm{x},\Lambda}(\omega)&=&-\frac{\mathrm{i}}{2\pi}\sum_{22'}\int d\Omega S_{22'}^{\Lambda}(\Omega)\gamma_{1'2'12}^{\mathrm{x},\Lambda}(\Omega-\omega),\nonumber\\
\partial_{\Lambda}\Sigma_{1'1}^{\mathrm{d},\Lambda}(\omega)&=&-\frac{\mathrm{i}}{2\pi}\sum_{22'}\bar{\gamma}_{1'2'12}^{\mathrm{d},\Lambda}(0)\int d\Omega S_{22'}^{\Lambda}(\Omega).
\end{eqnarray}
Note that the last term in the $\mathrm{d}$ channel includes the initial bare vertex $\bar{v}$.

In order to set up flow equations for the different channels, one assumes that the flow of $\mathrm{p}$, $\mathrm{x}$ and $\mathrm{d}$ is governed by the first, second and third terms on the RHS of Eq.~\eqref{eq:sord} and that there is no inter-channel coupling, i.e., each channel only enters its own flow equation. This yields
\begin{equation}
    \label{eq:chan_decomp_flow}
    \begin{split}
        \partial_\Lambda \gamma^{\mathrm{p},\Lambda}_{1'2'12}(\Pi)=\frac{\mathrm{i}}{2\pi}\int d\Omega\sum_{33'44'}\bar\gamma^{\mathrm{p},\Lambda}_{1'2'34}\left(\Pi\right) & S^\Lambda_{33'}\left(\frac{\Pi}{2}-\Omega\right)G^\Lambda_{44'}\left(\frac{\Pi}{2}+\Omega\right)\bar\gamma^{\mathrm{p},\Lambda}_{3'4'12}\left(\Pi\right),\\
        \partial_\Lambda \gamma^{\mathrm{x},\Lambda}_{1'2'12}(X)=\frac{\mathrm{i}}{2\pi}\int d\Omega\sum_{33'44'}\bar\gamma^{\mathrm{x},\Lambda}_{1'4'32}\left( X\right)\biggl[& S^\Lambda_{33'}\left(\Omega-\frac{X}{2}\right)G^\Lambda_{44'}\left(\Omega+\frac{X}{2}\right)\\
        + & G^\Lambda_{33'}\left(\Omega-\frac{X}{2}\right)S^\Lambda_{44'}\left(\Omega+\frac{X}{2}\right)\biggr]\bar\gamma^{\mathrm{x},\Lambda}_{3'2'14}\left(X\right),\\
        \partial_\Lambda \gamma^{\mathrm{d},\Lambda}_{1'2'12}(\Delta)=\frac{-\mathrm{i}}{2\pi}\int d\Omega\sum_{33'44'}\bar\gamma^{\mathrm{d},\Lambda}_{1'3'14}\left(\Delta\right)\biggl[& S^\Lambda_{33'}\left(\Omega-\frac{\Delta}{2}\right)G^\Lambda_{44'}\left(\Omega+\frac{\Delta}{2}\right)\\
        +& G^\Lambda_{33'}\left(\Omega-\frac{\Delta}{2}\right)S^\Lambda_{44'}\left(\Omega+\frac{\Delta}{2}\right)\biggr]\bar\gamma^{\mathrm{d},\Lambda}_{4'2'32}\left(\Delta\right).
    \end{split}
\end{equation}
We again emphasize that this is correct up to second order in the bare interaction. The initial condition given by Eq.~\eqref{eq:initial_conds} now reads $\gamma^{\{\mathrm{p},\mathrm{x},\mathrm{d}\},\Lambda\to\infty}_{1'2'12}=0$. Note that due to Eq.~(\ref{eq:sigma_channel_dec}), $\gamma^{\mathrm{d},\Lambda}$ is only needed at zero frequency.

The channel decomposition simplifies the flow equations considerably: the two-particle vertex is now defined over a single frequency grid, instead of the initial 3D grid space of the original flow equation. In addition, the multi-index structure simplifies:
\begin{equation}\label{eq:spi_support}
\begin{split}
\gamma^{\mathrm{p},\Lambda}_{1'2'12}(\Pi) &= 0 \hspace{10pt}\forall \hspace{2pt}\bar{v}_{1'2'12}\hspace{1pt}|\hspace{3pt}\bar{v}_{1'2'\bullet\bullet}=0\lor\bar{v}_{\bullet\bullet 12}=0,\\
\gamma^{\mathrm{x},\Lambda}_{1'2'12}(X) &= 0 \hspace{10pt}\forall \hspace{2pt}\bar{v}_{1'2'12}\hspace{1pt}|\hspace{3pt}\bar{v}_{1'\bullet\bullet 2}=0\lor\bar{v}_{\bullet2'1\bullet}=0,\\
\gamma^{\mathrm{d},\Lambda}_{1'2'12}(\Delta) &= 0 \hspace{10pt}\forall \hspace{2pt}\bar{v}_{1'2'12}\hspace{1pt}|\hspace{3pt}\bar{v}_{1'\bullet 1\bullet}=0\lor\bar{v}_{\bullet 2'\bullet 2}=0.
\end{split}
\end{equation}
The bullets indicate all possible multi-indices within a fixed pair of multi-indices, e.g., the term $\bar{v}_{1'2'\bullet\bullet}$ keeps the multi-indices $1',2'$ fixed, but encodes all multi-indices values in the third and fourth positions. Eq.~\eqref{eq:spi_support} is a direct consequence of Eq.~(\ref{eq:bar_channels}), $\gamma^{\{\mathrm{p},\mathrm{x},\mathrm{d}\},\Lambda\to\infty}_{1'2'12}=0$, and the fact that the flow equations \eqref{eq:chan_decomp_flow} preserve this structure. For interactions satisfying Eq.~\eqref{eq:r_params_h_v}, Eq.~\eqref{eq:spi_support} translates to:
\begin{equation}\label{eq:spi_finite_range}
\begin{split}
\gamma^{\mathrm{p},\Lambda}_{1'2'12}(\Pi) &= 0 \hspace{10pt}\forall |i_{1'}-i_{2'}|\geq R_{v} \lor |i_{1}-i_{2}|\geq R_{v},\\
\gamma^{\mathrm{x},\Lambda}_{1'2'12}(X) &= 0 \hspace{10pt}\forall |i_{1'}-i_{2}|\geq R_{v} \lor |i_{2'}-i_{1}|\geq R_{v},\\
\gamma^{\mathrm{d},\Lambda}_{1'2'12}(\Delta) &= 0 \hspace{10pt}\forall |i_{1'}-i_{1}|\geq R_{v} \lor |i_{2'}-i_{2
}|\geq R_{v}.
\end{split}
\end{equation} 
Eq.~\eqref{eq:spi_finite_range} simplifies the structure of the two-particle vertex tensor, which becomes sparse in this case; it also restricts two of the multi-index sums on the RHS of Eq.~(\ref{eq:chan_decomp_flow}). It should be emphasized that Eq.~\eqref{eq:spi_support} and (\ref{eq:spi_finite_range}) do not constitute any additional approximation; they are a consequence of the finite-ranged nature of the bare interaction vertex $\bar{v}_{1'2'12}$ in the single particle indices and the channel decomposition in Eq.~\eqref{eq:channel_decomp_pxd}. The numerical cost of evaluating the flow equations for $\gamma^\la$ at a given value of $\Lambda$ is then given by
\begin{equation}
\underbrace{O(L^2N_\Omega)}_{\text{\#components}}~~
\underbrace{O(L^2N_\Omega)}_{\text{sum/int on RHS}},
\end{equation}
where $N_\Omega$ denotes the number of points of the frequency grid, and $L$ parametrizes the number of single-particle degrees of freedom. The effort to determine $G^\Lambda$ is not included, it scales as $O(L^3N_\Omega)$.

\subsection{Expressing the flow equations in terms of convolutions}
\label{subsec:flow_eq_convos}

If we rewrite the RHS of the flow equations in terms of convolutions, we can make use of efficient convolution algorithms like the Fast Fourier Transform (FFT). The specific choice on how to perform the convolutions will depend on the problem at hand, and it will be affected by several factors like the frequency discretization and general performance requirements; here we will present the general description~\cite{Klockner_NJP}.

The convolution between two functions is defined as:
\begin{equation}\label{eq:conv_def}
    (f* g)(y)=\int dx f(x)g(y-x).
\end{equation}
The self-energy flow equations expressed in terms of convolutions read:
\begin{equation}
    \label{eq:flow_convo}
    \begin{split}
     \partial_\Lambda \Sigma^{\mathrm{p},\Lambda}_{1'1}(\omega)&=-\frac{\mathrm{i}}{2\pi}\sum_{22'}\tilde S^\Lambda_{22'}*\gamma^{\mathrm{p},\Lambda}_{1'2'12},\\
        \partial_\Lambda \Sigma^{\mathrm{x},\Lambda}_{1'1}(-\omega)&=-\frac{\mathrm{i}}{2\pi}\sum_{22'}\tilde S^\Lambda_{22'}*\gamma^{\mathrm{x},\Lambda}_{1'2'12},\\
        \partial_\Lambda \Sigma^{\mathrm{d},\Lambda}_{1'1}(\omega)&=-\frac{\mathrm{i}}{2\pi}\sum_{22'}\bar \gamma^{\mathrm{d},\Lambda}_{1'2'12}(0)\int d\Omega S^\Lambda_{22'}(\Omega),        
    \end{split}
\end{equation}
and the vertex flow equation are given by
\begin{equation}
    \label{eq:flow_convoB}
    \begin{split}
        \partial_\Lambda \gamma^{\mathrm{p},\Lambda}_{1'2'12}(\Pi)&=\frac{\mathrm{i}}{2\pi}\sum_{33'44'}
                                                       \bar\gamma^{\mathrm{p},\Lambda}_{1'2'34}\left(\Pi\right)\left[G^\Lambda_{44'}*S^\Lambda_{33'}\right](\Pi)~\bar\gamma^{\mathrm{p},\Lambda}_{3'4'12}\left(\Pi\right),\\
        \partial_\Lambda \gamma^{\mathrm{x},\Lambda}_{1'2'12}(X)&=\frac{\mathrm{i}}{2\pi}\sum_{33'44'}
        \bar\gamma^{\mathrm{x},\Lambda}_{1'4'32}\left( X\right)\biggl[G^\Lambda_{44'}*\tilde S^\Lambda_{33'} +S^\Lambda_{44'}*\tilde G^\Lambda_{33'}\biggr](X)~\bar\gamma^{\mathrm{x},\Lambda}_{3'2'14}\left(X\right),\\
        \partial_\Lambda \gamma^{\mathrm{d},\Lambda}_{1'2'12}(\Delta)&=\frac{-\mathrm{i}}{2\pi}\sum_{33'44'}\bar\gamma^{\mathrm{d},\Lambda}_{1'3'14}\left(\Delta\right)\biggl[G^\Lambda_{44'}*\tilde S^\Lambda_{33'}+ S^\Lambda_{44'}*\tilde G^\Lambda_{33'}\biggr](\Delta)~\bar\gamma^{\mathrm{d},\Lambda}_{4'2'32}\left(\Delta\right),        
    \end{split}
\end{equation}
where we have defined $\tilde{S}_{33'}^{\Lambda}(\omega)= S^{\Lambda}_{33'}(-\omega)$ and similarly for $\tilde{G}^{\Lambda}_{33'}$. These equations are local in the frequencies $\Pi,X,\Delta,\omega$; this implies that for the $\mathrm{d}$ channel, only components at the $\Delta=0$ point need to be obtained, since these are the only ones coupling to the self-energy flow equation. The above representation of the flow equations shows that the self-energy and the two-particle vertex functions are continuous functions of their corresponding frequencies. Since $S^{\Lambda}(\omega\to\pm\infty)\to 0$ and $G^{\Lambda}(\omega\to\pm\infty)\to 0$ quickly, any convolution term occurring on the RHS yields a continuous function.

\section{INFINITE, TRANSLATION-INVARIANT SYSTEMS: FRG APPROACH}\label{sec:app_inf}
We now apply the general second-order FRG scheme to an infinite system with a unit cell of size $L$ featuring translation invariance as specified in Sec.~\ref{subsubsec:inf_sys}. We follow Ref. \onlinecite{Klockner_NJP} where additional details can be found. The single-particle Hamiltonian, two-particle interactions, and hybridization functions are assumed to be of finite range, with corresponding range parameters $R_{h}$, $R_{v}$ and $R_{\Gamma}$ as described in Eqs.~\eqref{eq:r_params_h_v} and~\eqref{eq:finite_range_gamma}. The starting point are the flow equations (\ref{eq:ford2}) and (\ref{eq:chan_decomp_flow}).

It is reasonable to choose the FRG cut-off scheme such that the translational symmetry is respected. For this reason, the Fermi distribution functions $n^{\text{cut}}_{i}(\omega)$ corresponding to the set of artificial reservoirs must satisfy:
\begin{eqnarray}\label{eq:ni_cutoff}
n^{\text{cut}}_{(i+L)}(\omega)&=& n^{\text{cut}}_{i}(\omega-LE).
\end{eqnarray}
Due to the infinite system size, one needs to devise further approximations, which we will discuss in Sec.~\ref{subsec:inf_finite_support}.

\subsection{Exploiting the system's translational symmetry}\label{subsec:symmetries}
The translational symmetries of the system are given by Eqs.~(\ref{h_E_field}), (\ref{hyb_trans_inv}), and (\ref{sym_gsigma}), respectively. They are inherited by the single-scale propagator:
\begin{eqnarray}\label{eq:single_scale_prop_sym}
S^{\Lambda}_{(1'+L)(1+L)}(\omega)=S^{\Lambda}_{1'1}(\omega-LE).
\end{eqnarray}
For the two-particle vertex, one obtains
\begin{equation}\label{eq:shift_sym_vertex}
    \begin{split}
        \gamma^{\mathrm{p},\Lambda}_{(1'+L)(2'+L)(1+L)(2+L)}(\Pi)& = \gamma^{\mathrm{p},\Lambda}_{1'2'12}(\Pi-2LE),\\
        \gamma^{\mathrm{x},\Lambda}_{(1'+L)(2'+L)(1+L)(2+L)}(X)& = \gamma^{\mathrm{x},\Lambda}_{1'2'12}(X),\\
        \gamma^{\mathrm{d},\Lambda}_{(1'+L)(2'+L)(1+L)(2+L)}(\Delta)& = \gamma^{\mathrm{d},\Lambda}_{1'2'12}(\Delta).
    \end{split}
\end{equation}
These relations trivially hold for $\Lambda\to\infty$, and they are preserved by the flow equation. For the p channel, this can be seen as follows:
\begin{eqnarray}
\partial_\Lambda\gamma^{\mathrm{p},\Lambda}_{(1'+L)(2'+L)(1+L)(2+L)}(\Pi)	&=&\frac{\mathrm{i}}{2\pi}\sum_{33'44'}\int d\Omega \bar{\gamma}^{\mathrm{p},\Lambda}_{(1'+L)(2'+L)34}(\Pi)\biggl[ S_{33'}^{\Lambda}\left(\frac{\Pi}{2}-\Omega\right) G^{\Lambda}_{44'}\left(\frac{\Pi}{2}+\Omega\right)\biggr]\bar{\gamma}^{\mathrm{p},\Lambda}_{3'4'(1+L)(2+L)}(\Pi)\nonumber\\
&=&\frac{\mathrm{i}}{2\pi}\sum_{33'44'}\int d\Omega \bar{\gamma}^{\mathrm{p},\Lambda}_{1'2'34}(\Pi_-)\biggl[ S_{(3+L)(3'+L)}^{\Lambda}\left(\frac{\Pi}{2}-\Omega\right) G^{\Lambda}_{(4+L)(4'+L)}\left(\frac{\Pi}{2}+\Omega\right)\biggr]\bar{\gamma}^{\mathrm{p},\Lambda}_{3'4'12}(\Pi_-)\nonumber\\
&=&\frac{\mathrm{i}}{2\pi}\sum_{33'44'}\int d\Omega \bar{\gamma}^{\mathrm{p},\Lambda}_{1'2'34}(\Pi_-)\biggl[ S_{33'}^{\Lambda}\left(\frac{\Pi_-}{2}-\Omega\right) G^{\Lambda}_{44'}\left(\frac{\Pi_-}{2}+\Omega\right)\biggr]\bar{\gamma}^{\mathrm{p},\Lambda}_{3'4'12}(\Pi_-)\nonumber\\&=&
\partial_\Lambda\gamma^{\mathrm{p},\Lambda}_{1'2'12}(\Pi-2LE),
\end{eqnarray}
where $\Pi_-=\Pi-2LE$. The argument for the x and d channels is similar, but the frequency shift in the last step can be absorbed in the integration variable. The relations (\ref{eq:shift_sym_vertex}) can now be inserted into the self-energy flow equation (\ref{eq:ford2}) to show that the symmetries in Eq.~(\ref{sym_gsigma}) also hold within the FRG approximation:
\begin{equation}
\begin{split}
    \partial_\Lambda\Sigma^\Lambda_{(1'+L)(1+L)}(\omega) &=-\frac{\mathrm{i}}{2\pi} \int d\Omega\sum_{22'}  S^\Lambda_{(2+L)(2'+L)}(\Omega)
    \Big[\bar v_{1'2'12}+ \gamma^{\mathrm{p},\Lambda}_{1'2'12}(\Omega+\omega-2LE)+\gamma^{\mathrm{x},\Lambda}_{1'2'12}(\Omega-\omega)+\gamma^{\mathrm{d},\Lambda}_{1'2'12}(0)\Big] \\
&=-\frac{\mathrm{i}}{2\pi} \int d\Omega\sum_{22'}  S^\Lambda_{22'}(\Omega)
    \Big[\bar v_{1'2'12}+ \gamma^{\mathrm{p},\Lambda}_{1'2'12}(\Omega+\omega-LE)+\gamma^{\mathrm{x},\Lambda}_{1'2'12}(\Omega+LE-\omega)+\gamma^{\mathrm{d},\Lambda}_{1'2'12}(0)\Big]\\
& = \partial_\Lambda\Sigma^\Lambda_{1'1}(\omega-LE).
\end{split}\end{equation}
When solving the flow equations, one of the spatial indices can thus be restricted to be within the unit cell:
\begin{eqnarray}\label{eq:ind_comp_vertex}
\Sigma^{\{\mathrm{p,x,d}\},\Lambda}_{1'1}:~i_{1'}\in\{0,...,L-1\},~~~~~
\gamma^{\{\mathrm{p,x,d}\},\Lambda}_{1'2'12}:~i_{1'}\in\{0,...,L-1\}.
\end{eqnarray} 
The elements of this set are the ones that are explicitly calculated when solving the flow equations.

\subsection{Further approximation: finite support of the vertex functions}\label{subsec:inf_finite_support}
As mentioned above, further approximations to the flow equations need to be devised for an infinite system. The key idea is to limit the spatial support of the vertex functions, which is physically motivated by the fact that inelastic scattering will in general be limited by the system's correlation length. To this end, we choose a number $M\in\mathbb{N}$, $M\geq R_{v}$ and set~\cite{Klockner_NJP}:
\begin{alignat}{3}
    \label{eq:truncM}
    \Sigma_{1'1}^\Lambda&=0\hspace{2mm}&&\forall &|i_{1'}-i_{1}|&\geq M, \nonumber\\
    \gamma_{1'2'12}^\Lambda&=0\ &&\forall\ &\text{dist}(i_{1'},i_{2'},i_{1},i_{2})&\geq M.
\end{alignat} 
This limits the number of non-zero components of both $\Sigma^\Lambda$ and $\Gamma^\Lambda$ and restricts the single-particle sums on the RHS of the flow equations (\ref{eq:ford2}) and (\ref{eq:chan_decomp_flow}). In the $M\to \infty$ limit, the original flow equations (\ref{eq:ford2}) and (\ref{eq:chan_decomp_flow}) are recovered. $M$ becomes a numerical control parameter of the ODEs system, and observables should always reach convergence with respect to $M$. In combination, Eqs.~(\ref{eq:ind_comp_vertex}) and (\ref{eq:truncM}) entail that the numerical cost of evaluating the flow equations for $\gamma^\la$ at a given value of $\Lambda$ now reads
\begin{equation}
\underbrace{O(LMN_\Omega)}_{\text{\#components}}~~
\underbrace{O(M^2N_\Omega)}_{\text{sum/int on RHS}},
\end{equation}
which does not include the computation of $G^\Lambda$. To allow for a numerical solution of the flow equations in  Eqs.~\eqref{eq:flow_convo} and \eqref{eq:flow_convoB}, the frequency dependence of the vertex functions is discretized on a grid of $N_{\Omega}$ points. Details on this can be found in Ref.~\onlinecite{Klockner_NJP}.

In general, the Green's function $G^\Lambda$ and the single-scale propagator $S^\Lambda$ that enter the RHS of the flow equations cannot be computed for an infinite system. In our case, however, Eq.~\eqref{eq:truncM} implies that only a limited number of components contribute to the RHS. Within the self-energy flow equation (\ref{eq:ford2}), the distance of the single-particle indices is limited to $|i_{2}-i_{2'}|<M$. Likewise, the inner sum of spatial indices within the two-particle vertex flow equations (\ref{eq:chan_decomp_flow}) will be restricted to $\text{dist}(i_{3'},i_{4'},i_{3},i_{4})<3M$. Thus, $G^\Lambda$ and $S^\Lambda$ are only needed within a limited region, and we can employ the techniques of Sec.~\ref{subsec:iter_gf} by setting $N=3M$.

\subsection{Iterative techniques for the single-scale propagators }\label{subsec:iterative_s}

In Sec.~\ref{subsec:iter_gf} we showed how the Green's functions $G^\Lambda$ of an infinite system can be computed iteratively. Since the \text{RHS} of the flow equations also depends on the single-scale propagators $S^{\Lambda}(\omega)$, here we show how to calculate these quantities by means of the same iterative methods~\cite{Klockner_NJP}. Note that for $S^{\Lambda}(\omega)$, the choice of the cut-off affects the terms appearing in the expressions due to the definition in Eq.~\eqref{eq:single_scale_prop}. We frequently omit writing out the $\Lambda$-dependence.

We begin with the retarded part of the single-scale propagator; the advanced part follows from $[S^\tn{ret}]^\dagger = S^\tn{adv}$. In the reservoir cut-off scheme, the flow parameter $\Lambda$ only enters in the diagonal terms of $[G^\tn{ret}]^{-1}$ in Eq.~\eqref{block_structure_gret} and thus
\begin{equation}
    \begin{split}
    \pa D = \mathrm{i},~~
        \partial_\Lambda^* T_{\ci\li}=\partial_\Lambda^* T_{\li\ci}=\partial_\Lambda^* T_{\ci\ri}=\partial_\Lambda^* T_{\ri\ci}=0.
    \end{split}
\end{equation}
The following treatment allows a straightforward generalization for more general cut-off schemes.

Taking the derivative of Eq.~(\ref{eq:iter_greta}), the retarded part of the single-scale propagator reads:
\begin{equation}
\begin{split}
 S^\tn{ret}_{\ci\ci} = \partial_\Lambda^* G^\tn{ret}_{\ci\ci} = - G^\tn{ret}_{\ci\ci}\Big[ 
\mathrm{i}  &- T_{\ci\li} \calSrLR_{\li\li}T_{\li\ci}
 - T_{\ci\ri} \calSrLR_{\ri\ri}T_{\ri\ci}
\Big] G^\tn{ret}_{\ci\ci}.
\end{split}
\end{equation}
The auxiliary terms appearing in this equation are $\calSrLR=\pa\calrLR$; they can be obtained by taking the derivative in Eq.~(\ref{eq:iter_gf2}):
\begin{equation}\label{eq:iter_sr2}
\begin{split}
&~ \calSrLR_{\li\li}(\omega-NE)  =-\calrR_{\ci\ci}(\omega)\Big[\mathrm{i} - T_{\ci\li}\calSrLR_{\li\li}(\omega)T_{\li\ci}\Big]\calrR_{\ci\ci}(\omega),
\end{split}
\end{equation}
and similarly for $\calSrLR_{\ri\ri}$. This equation can be solved locally in a self-consistent way when \(E=0\), or successively using the boundary conditions \(\calSrLR(\pm\infty)=0\) for the case \(E\neq0\).

The Keldysh components of the single-scale propagator $S^{\K,\Lambda}$ follow in a similar way. The cutoff only enters into the diagonal of $\tsk$, which leads to [see Eqs.~\eqref{hybridizations}, ~\eqref{eq:tsk}, and \eqref{eq:cut_off_hyb}]
\begin{equation}
    \begin{split}
        \partial_\Lambda^* \tsk_{ij}(\omega)& =\partial_\Lambda \Gamma^{\mathrm{K},\Lambda}_{ij}(\omega)
        =-2\mathrm{i}[1-2n^{\text{cut}}_{i}(\omega)]\delta_{ij}. \\        
    \end{split}
\end{equation}
Taking the derivative of Eq.~(\ref{eq:iter_gk1}) yields (each of the terms carries a frequency argument $\omega$, which has been omitted for notation purposes)
\begin{equation}
\begin{split}
 S^\tn{K}_{\ci\ci}(\omega) = S^\tn{ret}_{\ci\ci}(\omega)\Big[\cdots\Big]G^\tn{adv}_{\ci\ci}(\omega)+&G^\tn{ret}_{\ci\ci}(\omega)\Big[\cdots\Big]S^\tn{adv}_{\ci\ci}(\omega) \\
+ G^\tn{ret}_{\ci\ci}\Big[ \pa\tsk_{\ci\ci} - & T_{\ci\li}\calSrLR_{\li\li}\tsk_{\li\ci} - \tsk_{\ci\li} \calSaLR_{\li\li} T^\dagger_{\li\ci} \\
 -&T_{\ci\ri}\calSrLR_{\ri\ri}\tsk_{\ri\ci}  - \tsk_{\ci\ri} \calSaLR_{\ri\ri} T^\dagger_{\ri\ci} 
  + T_{\ci\li}\calSkLR_{\li\li} T^\dagger_{\li\ci}+ T_{\ci\ri}\calSkLR_{\ri\ri}T^\dagger_{\ri\ci}\Big]G^\tn{adv}_{\ci\ci},
\end{split}
\end{equation}
where the brackets $[\ldots]$ are identical to the bracket in Eq.~(\ref{eq:iter_gk1}). To solve this equation, it remains to obtain the unknown quantities $\calSkLR_{\li\li}=\pa\calkLR_{\li\li}$; they are obtained by taking the derivative of Eq.~(\ref{eq:iter_gk2}):
\begin{equation}\label{eq:iter_sk2}
\begin{split}
&~ \calSkLR_{\li\li}(\omega-NE) \\= &~\calSrR_{\ci\ci}(\omega)\Big[\cdots\Big]\calaR_{\ci\ci}(\omega)+\calrR_{\ci\ci}(\omega)\Big[\cdots\Big]\calSaR_{\ci\ci}(\omega) \\
+ &~ \calrR_{\ci\ci}\Big[ \pa\tsk_{\ci\ci} -  T_{\ci\li}\calSrLR_{\li\li}\tsk_{\li\ci} - \tsk_{\ci\li} \calSaLR_{\li\li} T^\dagger_{\li\ci} 
  + T_{\ci\li}\calSkLR_{\li\li} T^\dagger_{\li\ci}\Big]\calaR_{\ci\ci},
\end{split}
\end{equation}
where the brackets $[\ldots]$ are identical to the bracket in Eq.~(\ref{eq:iter_gk2}), and the frequency argument $\omega$ for all quantities has been omitted in the last two lines. Note that Eq.~(\ref{eq:iter_translation2}) implies $\calSrR_{\ci\ci}(\omega)=\calSrLR_{\li\li}(\omega-NE)$; this quantity was obtained in a previous step via Eq.~(\ref{eq:iter_sr2}). Eq.~(\ref{eq:iter_sk2}) is an auxiliary equation for the Keldysh single-scale propagators, which becomes local when $E=0$ and can be solved self-consistently. For the more general case $E\neq0$, the equation can be solved iteratively~\cite{Klockner_NJP}.

\section{INFINITE, TRANSLATION-INVARIANT SYSTEMS: RESULTS}
\label{sec:inf_sys_res}

The methodology developed in the last section is now applied to study the non-equilibrium phase diagram of an interacting Wannier-Stark ladder coupled to reservoirs. We follow Refs.~\onlinecite{Klockner_PRL} \& \onlinecite{Klockner_NJP}, where additional details can be found.

\subsection{Model} 
\label{subsec:phys_model}

\begin{figure}[!t]
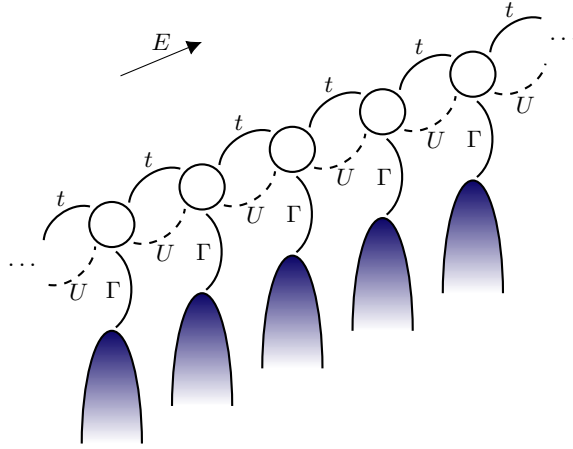

    \syst{-0.5}{1}
    \caption{(From Ref.~\onlinecite{Klockner_NJP}.) Sketch of the model used in Sec.~\ref{subsec:phys_model}. A tight-binding chain with hoppings \(t\) and interactions \(U\) is subject to an electrical field $E$ and is coupled to wide-band reservoirs with a hybridization \(\Gamma\).}\label{fig:pic_chain}
\end{figure}

The Wannier-Stark Hamiltonian is given by (see Fig.~\ref{fig:pic_chain})
\begin{eqnarray}\label{eq:H_inf}
H_{\text{sys}} &=& t\sum_{j\in\mathbb{Z}}\left(c_{j}^{\dagger}c_{j+1} + \text{H.c.}\right) + U\sum_{j\in\mathbb{Z}} \left( c_{j}^{\dagger}c_{j}-\frac{1}{2}\right)\left(c_{j+1}^{\dagger}c_{j+1}-\frac{1}{2}\right) + \sum_{j\in\mathbb{Z}}\left[ s(-1)^{j} + jE\right]c_{j}^{\dagger}c_{j},
\end{eqnarray}
where $t$ is the hopping strength, $U$ is the interaction parameter associated with the antisymmetric vertex $v_{0101}=-v_{0110}=U$, $s$ plays the role of a staggered potential, and $E$ represents the electric field. Each site $j$ of the chain is coupled to a single wide-band zero-temperature reservoir with a distribution function
\begin{eqnarray}
n^\nu(\omega)=\theta(\mu_\nu-\omega),~~ \mu_0=0, ~~\mu_{\nu+L}=\mu_{\nu}+LE.
\end{eqnarray}
We choose the distribution function of the auxiliary cut-off reservoirs to be identical to the one of the physical reservoirs, and the hybridizations are thus given by [compare Eqs.~(\ref{eq:wide_band_limit}) and \eqref{eq:cut_off_hyb}]:
\begin{eqnarray}\label{eq:cutoff_hyb_infsys}
\Gamma_{ij}^{\text{ret},\Lambda}(\omega) =  -\mathrm{i}\delta_{ij}(\Gamma+\Lambda),~~~
\Gamma^{\K,\Lambda}_{ij}(\omega) = -2\mathrm{i}\left[1-2n^{\nu=i}(\omega)\right](\Gamma+\Lambda)\delta_{ij},
\end{eqnarray}
where $\Gamma$ is the strength of the reservoir coupling. All terms of the Hamiltonian follow the discrete translational symmetries presented in Sec.~\ref{subsubsec:inf_sys}. For $s=0$, the size of the unit-cell is $L=1$, and $E$ in Eq.~(\ref{eq:H_inf}) corresponds to the parameter $E$ of Sec.~\ref{subsubsec:inf_sys}; for $s\neq0$, we have $L=2$ and $E\leftrightarrow E/2$.

\subsection{Fundamental questions to adress}\label{subsec:questions_inf}

We briefly recapitulate known results about the phase diagram. The equilibrium point with $E=\Gamma=0$ but finite $U$ is exactly solvable by Bethe ansatz~\cite{Giamarchi2004}. For $U\leq 2t$, the model is a gapless Luttinger liquid, whereas for $U>2t$, translational symmetry is spontaneously broken, and the system is in a Mott-insulating phase with a degenerate ground state and charge-density wave (CDW) order. In the limit $U=\Gamma=0$, one recovers the non-interacting Wannier-Stark ladder where any finite $E$ localizes all the electron wavefunctions and the system becomes a Wannier-Stark insulator \cite{Wannier1962,Aoki2014,Davison1997,Neumayer2015}. A finite coupling to the environment $\Gamma> 0$ induces finite currents, and the spectral function features peaks which are broadened on a scale $\Gamma$ \cite{Han2013}.

Along the equilibrium line $E=0$ but for $U,\Gamma> 0$, one expects that $U$ and $\Gamma$ have competing effects: While large $U$ favour CDW order, finite reservoir couplings $\Gamma$ tend to delocalize charge and favour metallic behaviour. At $\Gamma=0$ but for finite $U,E$, the system is in non-equilibrium, isolated, and interacting; this represents a scenario where Stark many-body localization effects might come into play~\cite{Schulz2019}. However, in this case the formation of a non-equilibrium steady state is under question, because a continuous application of $E$ will generically drive the system to an infinite temperature state. One expects that such heating is suppressed by a finite reservoir coupling $\Gamma$.

The generic non-equilibrium phase diagram ($E,\Gamma>0$) in the presence of interactions $U>0$ is not known and will now be computed using the novel FRG scheme; details can be found in Refs.~\onlinecite{Klockner_NJP,Klockner_PRL}.

\subsection{Benchmark of iterative Green's functions techniques}
\label{subsec:gf_benchmark}

The second-order FRG scheme presented in Sec.~\ref{sec:app_inf} strongly relies on the computation of the Green functions $G^{\Lambda}(\omega)$ and the single scale propagators $S^{\Lambda}(\omega)$ in an infinite system. This can be done using the iterative method discussed in Sec.~\ref{subsec:iter_gf}, and it is instructive to benchmark this technique in the exactly solvable case $U=0$.

\begin{figure}[t!]
    \includegraphics[width=0.3\linewidth,clip]{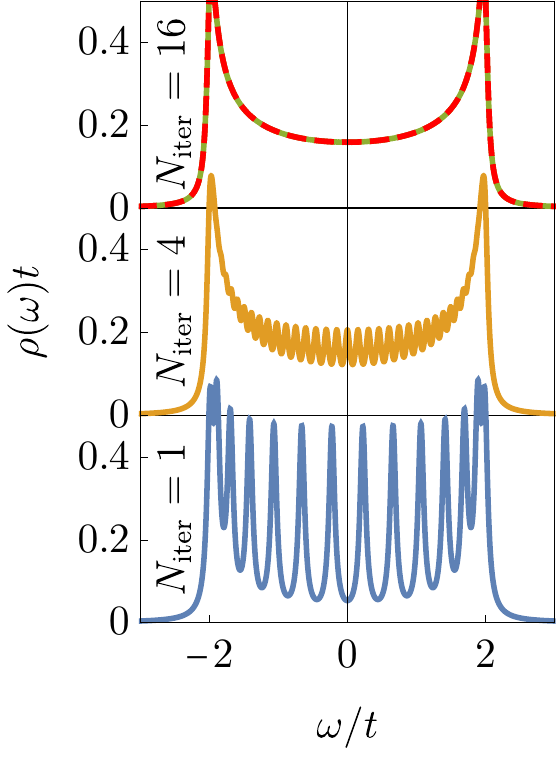}
    \caption{(From Ref.~\onlinecite{Klockner_NJP}.) 
        Benchmarking of the iterative, infinite-system Green's function algorithm introduced in Sec.~\ref{subsubsec:iter_gf1} in the non-interacting limit $U=0$. The local density of states at \(\Gamma/t=0.05\) and \(E=0\) is computed with \(1,\ 4,\ 16\) iterations of the self-consistency equation~\eqref{eq:iter_gf2} (blue, yellow and green; the curves are shifted vertically for readability).
        The dashed red line shows the analytical result in Eq.~(\ref{eq:dosexact}).
    }\label{fig:dos}
\end{figure}

For $U=E=s=0$, the local density of states is given by
\begin{eqnarray}\label{eq:rho_ldos}\label{eq:dosexact}
\rho(\omega) = -\frac{1}{\pi}\text{Im}\left(g_{ii}^{\text{ret}}(\omega)\right)=-\frac{1}{\pi}\text{Im}\left(\frac{1}{\sqrt{\left(\omega+\I\Gamma\right)^2-4t^2}}\right),
\end{eqnarray} 
which is independent of the site $i$. The algorithm of Sec.~\ref{subsubsec:iter_gf1} can be applied with $N=L=1$, and the analytical result can be reproduced easily by solving Eq.~\eqref{eq:iter_gf2} via a self-consistency loop (see Fig.~\ref{fig:dos}).

\subsection{FRG results for the non-equilibrium phase diagram}
\label{subsec:app_comp_mft}

\begin{figure}[t!]
    \begin{overpic}[scale=1.0,width=0.5\columnwidth]{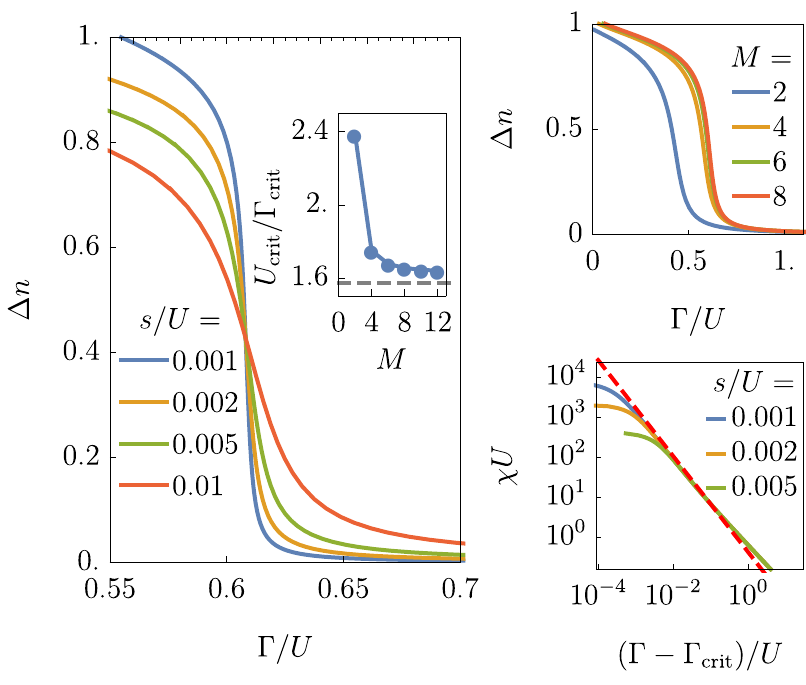}
        \put(60,78){(c)}
        \put(60,39){(d)}
        \put(0,78){(a)}
        \put(50,67){(b)}
    \end{overpic}
    \caption{ (From Ref.~\onlinecite{Klockner_NJP}.) Functional RG data for $t=0$ and comparison with mean-field results.
        (a) CDW order parameter $\Delta n$ as a function of $\Gamma/U$ for various values of $s$. The transition into the CDW phase becomes sharper as $s$ is decreased. The data was obtained for fixed $M=10$, which is the numerical control parameter in the solution of the flow equations.
        (b) \(M\)-dependence of \(U_\mathrm{crit}/\Gamma_\mathrm{crit}\), which is defined as the point in (a) where lines at different \(s\) intersect. The dashed line shows the mean-field result of Eq.~(\ref{eq:ucritmf}).
        (c) CDW order parameter as a function of $\Gamma$ for constant $s/U=0.01$ but various $M$; convergence in $M$ can be reached easily.
        (d) In agreement with mean-field data, $\chi$ exhibits a power-law divergence with an exponent \(\gamma\approx 1.2\) close to the critical point (dashed line).
    }\label{fig:large_inter_phase}
\end{figure}

We now use the FRG to compute the phase diagram for generic values of $E$, $\Gamma$, and $U$. Details can be found in Refs.~\onlinecite{Klockner_NJP,Klockner_PRL}. The CDW and metallic phases can be discerned via the charge susceptibility:
\begin{eqnarray}\label{eq:chi_sus}
\chi &=& \lim_{s\to0}\frac{\Delta n}{s},\hspace{6pt}\Delta n = \langle c_{2j}^{\dagger}c_{2j} - c_{2j+1}^{\dagger}c_{2j+1}\rangle,
\end{eqnarray}
which diverges in the CDW phase.

\subsubsection{Comparison with mean-field theory}
\label{subsubsec:mf_comparison}

We first study the limit $U/t\gg1$ which can be treated accurately using mean-field theory (our model is spinless and simply maps to an Ising chain). Since the FRG is strictly controlled only to second order in $U$, this represents a highly-non-trivial test. At $t=0$, the effects of the external electric field can be eliminated by means of a gauge transformation, and the mean-field equations take the form~\cite{Klockner_NJP}:
\begin{equation}\label{eq:mft_eq}
\Delta n=\frac{2}{\pi}\arctan\left( s/\Gamma+U\Delta n/\Gamma \right).
\end{equation} 
For $s\to0$, this equation only has CDW solutions with $\Delta n\neq 0$ beyond some critical interaction strength:
\begin{equation}\label{eq:ucritmf}
    U^\mathrm{MF}_\mathrm{crit}=\frac{\pi}{2}\Gamma^\mathrm{MF}_\mathrm{crit}\approx 1.571\Gamma^\mathrm{MF}_\mathrm{crit}.
\end{equation}
For $U/\Gamma <\pi/2$, the system is metallic, and the susceptibility scales as:
\begin{equation}\label{eq:chi_scaling}
  \chi^\mathrm{MF}=\lim_{s\to 0}\frac{\Delta n}{s}\sim \frac{1/\Gamma}{\pi/2-U/\Gamma},\hspace{0.5cm} U/\Gamma<\pi/2.
\end{equation}

In Fig.~\ref{fig:large_inter_phase}, we compare these mean-field results with FRG data for $t=0$. Panel (a) shows the order parameter $\Delta n$ as a function of $\Gamma/U$; the FRG correctly captures the phase transition. The critical interaction $U_{\text{crit}}/\Gamma_{\text{crit}}$ can be identified as the point where the curves for different $s$ intersect. The FRG prediction is in good agreement with the mean-field result; moreover, convergence in the control parameter $M$ (which governs the spatial extent of the vertex functions, see Sec.~\ref{subsec:inf_finite_support}) can easily be reached, see Fig.~\ref{fig:large_inter_phase}(b) and (c). The FRG also predicts a power-law divergence of the susceptibility $\chi$ around the critical region, see Fig.~\ref{fig:large_inter_phase}(d).

By construction, the FRG is exact in the non-interacting limit. We have just shown that it also accurately reproduces the mean-field results for $t/U=0$. This is a highly non-trivial result.  Note that in contrast to mean field, the FRG also correctly captures the metal-CDW transition at $E=\Gamma=0$ (see below and Ref.~\onlinecite{Klockner_PRL}).

\subsubsection{Generic phase diagram}

\begin{figure}[!t]
    \includegraphics[width=0.7\columnwidth,clip]{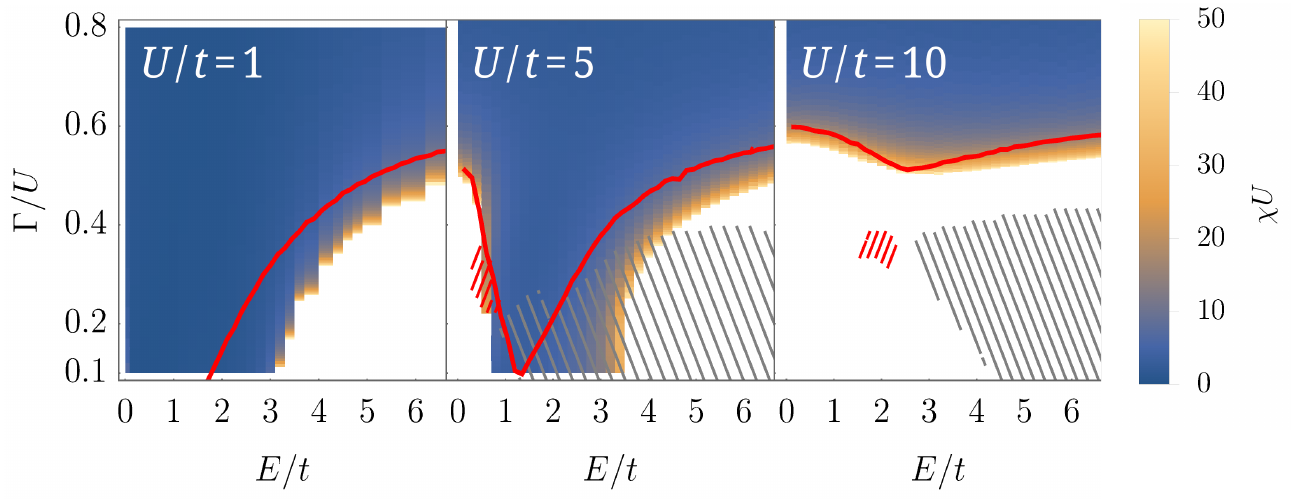}
    \caption{ (Adapted from Ref.~\onlinecite{Klockner_PRL}.) Phase diagram as a function of $\Gamma$ and $E$ for different values of
$U$. At large $U/t=10$, FRG data agrees well with mean-field results (red line). At intermediate $U/t=5$, FRG correctly captures the CDW phase for $\Gamma=E=0$. Application of an electric field first induces currents and drives the system into a metallic, but eventually many-body Wannier-Stark localization sets at large $E$ in and the system reenters a CDW phase. For small $U/t=1$, the transition into an CDW phase is driven solely by $E$. Grey and red hatching indicates regions where no unique and no stable mean-field solution exists, respectively (see Ref.~\onlinecite{Klockner_PRL} for details).}\label{fig:phase_diagram}
\end{figure}

We now discuss FRG data for the phase diagram for arbitrary values of the $\Gamma,E$ and $U$; results are shown in Fig.~\ref{fig:phase_diagram}, more details can be found in Ref.~\onlinecite{Klockner_PRL}. At large (but finite) $U/t=10$, the FRG again reproduces mean-field results accurately. Another benchmark (which is not shown in Fig.~\ref{fig:phase_diagram}; see Ref.~\onlinecite{Klockner_PRL}) corresponds to the case with $E=\Gamma=0$, where FRG yields a critical value of $U_{\text{crit}}/t\approx 1.4$ for the transition between the metallic and CDW phases in stark contrast with mean field, which gives $U_{\text{crit}}^{\text{MF}}=0$. The CDW phase for $E=\Gamma=0$ manifests in the lower-left corner of the phase diagram for $U/t=5$. Application of a finite electric field $E$ restores translational symmetry through induced currents and drives the system into a metallic phase. However, sufficiently strong values of $E$ eventually drive the system back to a CDW phase. At small $U/t=1$, it is solely the electric field that drives the transition into the CDW phase.

\section{FINITE SYSTEMS: FRG APPROACH}\label{sec:app_finite}

We turn our attention to the case of finite systems of size $L$. In this case, it is reasonable to choose a distribution function of the auxiliary reservoirs that is independent of the site $i$:
\begin{eqnarray}\label{eq:ni_cutoff2}
n^{\text{cut}}(\omega)= n^{\text{cut}}_{i}(\omega).
\end{eqnarray}
As before, the flow equations need to be simplified in order to tackle systems of more than a few sites. We summarize the approach of Ref.~\onlinecite{Klockner_PRB_2020}.

\subsection{Simplification of the flow equations}\label{subsec:finite_flow_eqns}

The flow equations in the channel decomposition [Eqs.~\eqref{eq:ford2} and~\eqref{eq:chan_decomp_flow}] can in practice not be solved for a large system. A straightforward simplification is to set $\gamma^{\Lambda}=\bar{v}$ on the $\text{RHS}$, which is equivalent to setting $\gamma^{\Lambda}=\bar{v}$ directly on the RHS of Eq.~\eqref{eq:sord}. For didactic purposes, we again state the resulting flow equations:
\begin{equation}\label{eq:se_ford2}
\begin{split}
    \partial_\Lambda&\Sigma^\Lambda_{1'1}(\omega) =-\frac{\mathrm{i}}{2\pi} \int d\Omega\sum_{22'}  S^\Lambda_{22'}(\Omega)\Big[\bar v_{1'2'12}+ \gamma^{\mathrm{p},\Lambda}_{1'2'12}(\Omega+\omega)+\gamma^{\mathrm{x},\Lambda}_{1'2'12}(\Omega-\omega)+\gamma^{\mathrm{d},\Lambda}_{1'2'12}(0)\Big],
\end{split}
\end{equation}
as well as
\begin{equation}
    \label{eq:chan_decomp_flow_with_bare_vertex}  
    \begin{split}  
        \partial_\Lambda \gamma^{\mathrm{p},\Lambda}_{1'2'12}(\Pi)=\frac{\mathrm{i}}{2\pi}\int d\Omega\sum_{33'44'}\bar v_{1'2'34}\Big[ & S^\Lambda_{33'}\left(\frac{\Pi}{2}-\Omega\right)G^\Lambda_{44'}\left(\frac{\Pi}{2}+\Omega\right)\Big]\bar v_{3'4'12},\\
        \partial_\Lambda \gamma^{\mathrm{x},\Lambda}_{1'2'12}(X)=\frac{\mathrm{i}}{2\pi}\int d\Omega\sum_{33'44'}\bar v_{1'4'32}\Big[& S^\Lambda_{33'}\left(\Omega-\frac{X}{2}\right)G^\Lambda_{44'}\left(\Omega+\frac{X}{2}\right)\\
        +& G^\Lambda_{33'}\left(\Omega-\frac{X}{2}\right)S^\Lambda_{44'}\left(\Omega+\frac{X}{2}\right)\Big]\bar v_{3'2'14},\\
        \partial_\Lambda \gamma^{\mathrm{d},\Lambda}_{1'2'12}(\Delta)=\frac{-\mathrm{i}}{2\pi}\int d\Omega\sum_{33'44'}\bar v_{1'3'14}\Big[& S^\Lambda_{33'}\left(\Omega-\frac{\Delta}{2}\right)G^\Lambda_{44'}\left(\Omega+\frac{\Delta}{2}\right)\\
        +& G^\Lambda_{33'}\left(\Omega-\frac{\Delta}{2}\right)S^\Lambda_{44'}\left(\Omega+\frac{\Delta}{2}\right)\Big]\bar v_{4'2'32}.   
        \end{split}
\end{equation}
A numerical implementation of Eq.~\eqref{eq:chan_decomp_flow_with_bare_vertex} is still too demanding due to the number of single-particle components. An additional approximation proposed in Ref.~\onlinecite{Klockner_PRB_2020} is to neglect the self-energy feedback in the two-particle vertex flow equations by replacing $G^{\Lambda}\to g^{\Lambda}$ and $S^{\Lambda}\to s^{\Lambda}$. The vertex flow equations can then be integrated; this is clear for the $\mathrm{x}$ and $\mathrm{d}$ channels whose RHS contains only $\partial_\Lambda g^\Lambda_{33'}g^\Lambda_{44'}$, and for the $\mathrm{p}$ channel, it follows from
\begin{eqnarray}\label{eq:p_channel_pert_deriv}
\partial_\Lambda \gamma^{\mathrm{p},\Lambda}_{1'2'12}(\Pi)&=&\frac{\mathrm{i}}{4\pi}\int d\Omega\sum_{33'44'}\left(\bar v_{1'2'34}\left[s^\Lambda_{33'}\left(\Omega_{-}\right)g^\Lambda_{44'}\left(\Omega_{+}\right)\right]\bar{v}_{3'4'12} + \bar{v}_{1'2'43}\left[s^\Lambda_{44'}\left(\Omega_{+}\right)g^\Lambda_{33'}\left(\Omega_{-}\right)\right]\bar{v}_{4'3'12} \right)\nonumber\\
&=&\frac{\mathrm{i}}{4\pi}\int d\Omega\sum_{33'44'}\bar v_{1'2'34}\left[s^\Lambda_{33'}(\Omega_{-})g^\Lambda_{44'}(\Omega_{+})+g^\Lambda_{33'}(\Omega_{-})s^\Lambda_{44'}(\Omega_{+})\right]\bar{v}_{3'4'12}\nonumber\\
&=&\partial_\Lambda\frac{\mathrm{i}}{4\pi}\int d\Omega\sum_{33'44'}\bar v_{1'2'34} g^\Lambda_{33'}g^\Lambda_{44'}\bar{v}_{3'4'12},~~~~
\Omega_{\pm} = \frac{\Pi}{2}\pm \Omega.
\end{eqnarray}
Restoring the frequency dependence on the RHS, the two-particle vertex flow equations in Eq.~\eqref{eq:chan_decomp_flow_with_bare_vertex} integrate to:
\begin{equation}
    \label{eq:chan_decomp_flow2}
    \begin{split}
        \gamma^{\mathrm{p},\Lambda}_{1'2'12}(\Pi)=&\frac{\mathrm{i}}{4\pi}\int d\Omega\sum_{33'44'}g^\Lambda_{33'}\left(\frac{\Pi}{2}-\Omega\right)g^\Lambda_{44'}\left(\frac{\Pi}{2}+\Omega\right)\bar v_{1'2'34}\bar v_{3'4'12},\\[1ex]
        \gamma^{\mathrm{x},\Lambda}_{1'2'12}(X)=&\frac{\mathrm{i}}{2\pi}\int d\Omega\sum_{33'44'}g^\Lambda_{33'}\left(\Omega-\frac{X}{2}\right)g^\Lambda_{44'}\left(\Omega+\frac{X}{2}\right)\bar v_{1'4'32}\bar v_{3'2'14},\\[1ex]
        \gamma^{\mathrm{d},\Lambda}_{1'2'12}(\Delta)=&\frac{-\mathrm{i}}{2\pi}\int d\Omega\sum_{33'44'}g^\Lambda_{33'}\left(\Omega-\frac{\Delta}{2}\right)g^\Lambda_{44'}\left(\Omega+\frac{\Delta}{2}\right)\bar v_{1'3'14}\bar v_{4'2'32}.
    \end{split}
\end{equation}
Eq.~\eqref{eq:chan_decomp_flow2} is nothing but the perturbation theory result for the two-particle vertex in presence of a reservoir coupling $\Lambda$. The absence of inelastic scattering contributions (i.e., neglecting the self-energy feedback) is numerically challenging when performing integrals, and in the next section we will show how the frequency integration on the RHS can be computed analytically. One should note that the self-energy flow equation \eqref{eq:se_ford2} is unaltered; the self-energy feeds back into its own flow equation through the dressed propagators $S^{\Lambda}$ appearing on the RHS.

The solution in Eq.~(\ref{eq:chan_decomp_flow2}) naturally satisfies Eqs.~(\ref{eq:spi_support}) and (\ref{eq:spi_finite_range}), and the components of $\gamma^{\{\mathrm{p},\mathrm{x},\mathrm{d}\},\Lambda}_{1'2'12}$ become sparse. If the interaction range $R_v$ is finite, the numerical cost of computing $\gamma^\la$ at a given $\Lambda$ scales as
\begin{equation}
\underbrace{O(L^2N_\Omega)}_{\text{\#components}}~~
\underbrace{O(N_\Omega)}_{\text{int on RHS}}.
\end{equation}
The effort to compute $g^\la$ is given by $O(L^3N_\Omega)$ and is not included.

\subsection{Analytic expressions for the two-particle vertex components}
\label{subsec:analytic_expressions}

We recap a semi-analytic approach to compute the two-particle vertex components that circumvents the numerical integration on the RHS \cite{Klockner_PRB_2020}. We define the modified single-particle Hamiltonian by:
\begin{equation}
    \bar h=h+\Gamma^{\tn{ret},\Lambda},
\end{equation}
where $\Gamma^{\tn{ret},\Lambda}$ is given in Eq.~\eqref{eq:cut_off_hyb}. For the case of wide-band reservoirs, this quantity is frequency independent by virtue of Eq.~\eqref{eq:wide_band_limit}. Note that \(\bar h\) is not hermitian, so it has separate left and right eigensystems:
\begin{equation}\label{eq:barheig}
    \begin{split}
        \bar h\left| q \right\rangle = \lambda_q  \left| q \right\rangle,~~~
        \left\langle \bar q \right| \bar h = \left\langle \bar q \right| \lambda_q.
    \end{split}
\end{equation}
The positivity of $\Gamma^{\tn{ret},\Lambda}$ ensures that $\text{Im}(\lambda_q)<0\ \forall q$. The retarded and advanced components of the non-interacting Green functions then read:
\begin{equation}
    \label{eq:grDecomp}
    \begin{split}
        g^{\tn{ret},\la}(\omega)&=\frac{1}{\omega-\bar h}=\sum_q \frac{1}{\omega-\lambda_q}\ket{q}\bra{\bar{q}}=\sum_q \frac{1}{\omega-\lambda_q}Q_q,\\
        g^{\tn{adv},\la}(\omega)&=\frac{1}{\omega-\bar h^\dagger}=\sum_q \frac{1}{\omega-\lambda^*_q}\ket{\bar q}\bra{q}=\sum_q \frac{1}{\omega-\lambda^*_q}Q^\dagger_q,
    \end{split}
\end{equation}
where we have introduced the matrix \(Q_q=\ket{q}\bra{\bar q}\). We now use Eq.~\eqref{eq:keldysh_sylvester_eq} as well as the Dyson equation (\ref{non_int_g}) to express $g^{\tn{K},\la}(\omega)$ in terms of an effective distribution function $n^\tn{eff}(\omega)$:
\begin{equation}\label{eq:sylvester2}\begin{split}
g^{\tn{K},\Lambda}& =g^{\tn{ret},\la}\Gamma^{\tn{K},\Lambda}g^{\tn{adv},\Lambda}= 
g^{\tn{ret},\la}(1-2n^\tn{eff}) - (1-2n^\tn{eff})g^{\tn{adv},\la}\\[1ex]
&\Leftrightarrow \Gamma^{\tn{K},\la} = \bar h (1-2n^\tn{eff}) - (1-2n^\tn{eff})\bar h^\dagger.
\end{split}\end{equation}
All reservoirs contribute additively to $n^\tn{eff}$, and the only frequency dependence stems from $n^\nu(\omega)$ and $n^\tn{cut}(\omega)$, which correspond to the distribution functions in the physical and auxiliary reservoirs, respectively [see Eq.~\eqref{eq:cut_off_hyb}]. As a result, the effective distribution function can be expressed in terms of some frequency-independent operators $\eta_\nu$ and $\eta_\tn{cut}$. In the special case that all reservoirs are at either zero or infinite temperature, one obtains:
\begin{equation}\begin{split}\label{eq:sylvesterIndiRes}
    1-2n^\tn{eff}(\omega) & = \sum_{\alpha=\nu,\tn{cut}} \eta_\alpha \left[1-2n^\alpha(\omega)\right] = \sum_{\substack{\alpha=\nu,\tn{cut}\\ T_\alpha=0}}\eta_\alpha\text{sgn}(\omega-\mu_\alpha).
\end{split}\end{equation}
By comparing Eqs.~(\ref{eq:restrRes}), (\ref{eq:sylvester2}), and (\ref{eq:sylvesterIndiRes}), matrix equations can be derived for $\eta_{\nu},\eta_{\text{cut}}$; they are given by
\begin{equation}
\begin{split}
    -2\mathrm{i}\Gamma^\nu&=\bar{h} \eta_\nu - \eta_\nu\bar h^\dagger,~~  -2\mathrm{i}\la=\bar h \eta_\tn{cut}- \eta_\tn{cut}\bar h^\dagger.
\end{split}
\end{equation}
These equations are of a Sylvester form, and they can be solved by the Bartels-Stewart algorithm \cite{Bartels1972} to obtain the matrices $\eta_{\nu}$ and $\eta_{\text{cut}}$, which in turn determine the effective distribution matrix $1-2n^{\text{eff}}$. Note that a unique solution for $n^\tn{eff}$ exists if and only if \(\bar h\) has no real eigenvalues, which is fullfilled when all degrees of freedom have a decay channel into one of the reservoirs.

Using Eqs.~(\ref{eq:grDecomp}) and (\ref{eq:sylvester2}), all terms appearing on the RHS of Eq.~(\ref{eq:chan_decomp_flow2}) can be expressed in terms of complex-valued integrals. As an specific example, let us consider the combination of $g^{\text{ret},\Lambda}_{i_{3}i_{3'}}g^{\K,\Lambda}_{i_{4}i_{4'}}$ on the RHS:
\begin{equation}
    \label{eq:exampleTwoGF}
    \begin{split}
    &\int \dOp \Omega\ g^{\tn{ret},\la}_{i_3i_{3'}}(\pm\Omega) g^{\tn{K},\la}_{i_4i_{4'}}(\Omega+\omega)\\
                                                                                  &=\int \dOp\Omega \sum_{q_1} \frac{1}{\pm\Omega-\lambda_{q_1}} \left(Q_{q_1}\right)_{i_3i_{3'}} \sum_{q_2}\sum_{\alpha}\sgn(\Omega+\omega-\mu_\alpha)
                                                                                   \biggl[\frac{1}{\Omega+\omega-\lambda_{q_2}} \left( Q_{q_2}\eta_\alpha\right)_{i_4i_{4'}}-\frac{1}{\Omega+\omega-\lambda_{q_2}^*} \left(\eta_\alpha Q^\dag_{q_2}\right)_{i_4i_{4'}}\biggr]\\[1ex]
                                                                                   &=\pm\sum_{q_1q_2}\sum_{\alpha} \left(Q_{q_1}\otimes Q_{q_2} \eta_\alpha\right)_{i_3i_{3'}i_4i_{4'}}f_1(\pm \lambda_{q_1}, \lambda_{q_2}-\omega,\mu_\alpha)
                                                                                   -\left(Q_{q_1}\otimes \eta_\alpha Q^\dag_{q_2} \right)_{i_3i_{3'}i_4i_{4'}}f_1(\pm \lambda_{q_1}, \lambda_{q_2}^*-\omega,\mu_\alpha),
\end{split}
\end{equation}
where $\omega\in\{\Pi,X,\Delta\}$. The integration variable $\Omega$ was shifted, and we introduced a frequency integral $f_1$
\begin{equation}
    \begin{split}
        f_1(a,b,\mu)&=\int d\Omega \frac{1}{\Omega-a} \frac{1}{\Omega-b} \sgn(\Omega-\mu).
    \end{split}
\end{equation}
This integral does not depend on the single-particle indices and can be computed analytically. All other terms can be treated similarly; the results can be found in Ref.~\onlinecite{Klockner_PRB_2020}.  The numerical effort to evaluate the vertex flow equation is now given by
\begin{equation}
\underbrace{O(L^2N_\Omega)}_{\text{\#components}}~~
\underbrace{O(L^2)}_{\text{sum on RHS}}.
\end{equation}
At each $\Lambda$, it is essential to first compute and store all quantities $Q_q$, $\eta_\alpha$, $Q_q\eta_\alpha$, $\eta_\alpha Q_q$. More details about the numerical implementation of the flow equations (in particular the issue of parallelization) can be found in Ref.~\onlinecite{Klockner_PRB_2020}.

\subsection{Perturbation theory limit}\label{subsec:pert_theory}

The expression for the two-particle vertex in Eq.~(\ref{eq:chan_decomp_flow2}) corresponds to second-order perturbation theory. We will now illustrate how the self-energy can be computed strictly up to second order in $U$~\cite{Klockner_PRB_2020}. This allows us to compute perturbation theory data using the existing numerics and to directly compare those with the FRG results.

The first-order contribution to the self-energy can be obtained by replacing the single-scale propagator as well as the two-body vertex on the RHS of Eq.~\eqref{eq:ford2} by their lowest-order expansion (\(s^\la\) and \(\bar v\), respectively) and by integrating the resulting equation \cite{Karrasch}:
\begin{equation}\label{eq:pt_first}
\begin{split}
    &\Sigma^{\tn{1PT},\la}_{1'1}(\omega) =-\frac{\I}{2\pi} \int d\Omega\sum_{22'}  g^\Lambda_{22'}(\Omega)\bar v_{1'2'12}.
\end{split}
\end{equation}
In order to compute the second-order term, we expand the single-scale propagator $S^{\Lambda}$:
\begin{equation}\label{eq:pt_s}
\begin{split}
S^\la&=\partial_\la^* G^\la =\partial_\la^*\left[g^\la+g^\la\Sigma^{\tn{1PT},\la}g^\la+\mathcal{O}(U^2)\right]  =s^\la+g^\la\Sigma^{\tn{1PT},\la} s^\la +s^\la\Sigma^{\tn{1PT},\la} g^\la+\mathcal{O}(U^{2})=s^\la +\mathcal{O}(U),
\end{split}
\end{equation}
where $s^{\Lambda}$ represents the single-scale propagator without self-energy feedback. Since \(\gamma^{\{\mathrm{p,x,d\}},\la}\sim U^2\), the second-order contribution to the RHS of Eq.~\eqref{eq:se_ford2} that is associated with the x- and p-channels is given by
\begin{equation}\label{eq:flow_pt_px}
\begin{split}
    &-\frac{\I}{2\pi} \int d\Omega\sum_{22'}  s^\Lambda_{22'}(\Omega)\Big[\gamma^{\mathrm{p},\Lambda}_{1'2'12}(\Omega+\omega)+\gamma^{\mathrm{x},\Lambda}_{1'2'12}(\Omega-\omega)\Big]    =-\partial_\Lambda \frac{\I}{2\pi} \int d\Omega\sum_{22'}  g^\Lambda_{22'}(\Omega)\Big[\gamma^{\mathrm{p},\Lambda}_{1'2'12}(\Omega+\omega)\Big].
\end{split}\end{equation}
The derivative in the last line acts on $g^\la$ as well as on $\gamma^{\tn{p},\la}$, which yields the first and second term in the first line, respectively. This follows from Eq.~(\ref{eq:chan_decomp_flow2}) by renaming indices as well as the integration variables. The second-order contributions to the RHS of Eq.~\eqref{eq:se_ford2} that are associated with the d-channel and with the single-scale propagator read:
\begin{equation}\label{eq:flow_pt_d}
\begin{split}
    &-\frac{\I}{2\pi} \int d\Omega\sum_{22'} s^\Lambda_{22'}(\Omega)\gamma^{\mathrm{d},\Lambda}_{1'2'12}(0)
     +\left\{s^\Lambda+g^\Lambda \Sigma^{\tn{1PT},\la} s^\Lambda+s^\Lambda \Sigma^{\tn{1PT},\la} g^\Lambda\right\}_{22'}(\Omega) \bar v_{1'2'12}\\
     =&-\partial_\Lambda\frac{\I}{2\pi} \int d\Omega\sum_{22'}  g^\Lambda_{22'}(\Omega)\Big[\bar v_{1'2'12} +\gamma^{\mathrm{d},\Lambda}_{1'2'12}(0)\Big],
\end{split}\end{equation}
where we have employed Eq.~(\ref{eq:pt_first}); the terms $g^\Lambda \Sigma^{\tn{1PT},\la} s^\Lambda\bar v$ can be identified with $g^\la \partial_\la\gamma^{\mathrm{d},\Lambda}$ via Eq.~(\ref{eq:chan_decomp_flow2}).

Eqs.~\eqref{eq:flow_pt_px} and \eqref{eq:flow_pt_d} constitute the RHS of the self-energy flow equation \eqref{eq:se_ford2} in second-order perturbation theory. Integrating w.r.t.~$\la$ gives:
\begin{equation}\label{eq:flow_pt}
\begin{split}
    \Sigma^{\tn{2PT},\la}_{1'1}(\omega) =-\frac{\I}{2\pi} \int d\Omega\sum_{22'}  g^\Lambda_{22'}(\Omega)\Big[\bar v_{1'2'12} + \gamma^{\mathrm{p},\Lambda}_{1'2'12}(\Omega+\omega)+\gamma^{\mathrm{d},\Lambda}_{1'2'12}(0)\Big].
\end{split}\end{equation}
Note that the RHS of the equation depends on the components $\gamma^{\tn{p},\Lambda}$ and $\gamma^{\tn{d},\Lambda}$, which are given in Eq.~\eqref{eq:chan_decomp_flow2}, and which can be computed analytically (see the last section); \(\gamma^{\mathrm{x},\Lambda}\) does not appear separately due to the combination of Eq.~\eqref{eq:flow_pt_px} with Eq.~\eqref{eq:chan_decomp_flow2}.  Eq.~(\ref{eq:flow_pt}) is completely general and is fullfilled even if there is a finite first-order contribution $\Sigma^{\tn{1PT},\la}$ to the self-energy \cite{Karrasch}.

\begin{figure*}
\begin{tikzpicture}
\coordinate(A0) at (0.6cm,0cm);
\foreach \a in {1,2,...,7}{
    \if\a4%
        \node[circle, minimum size=0.6cm, right=1.1cm*\a of A0, thick](A\a){$\dots$};
    \else
        \node[draw, circle, minimum size=0.6cm, right=1.1cm*\a of A0, thick](A\a){};
    \fi;
}
\foreach[evaluate=\a as \an using int(\a+1)] \a in {1,2,...,6}
    \path[]
        (A\a.north east) edge [
          text=black,
          thick,
          shorten <=2pt,
          shorten >=2pt,
          bend left=50
          ] node[above]{\(t\)} (A\an.north west);
\foreach[evaluate=\a as \an using int(\a+1)] \a in {1,2,...,6}
    \path[]
        (A\a.south east) edge [
          text=black,
          thick,
          shorten <=2pt,
          shorten >=2pt,
          bend right=50,
          dashed
          ] node[below]{\(U\)} (A\an.south west);

      \coordinate[left=2cm of A1](r1);
      \draw[thick, shading=axis, left color=white, right color=custBlue] (r1)++(-90:1.5 and 0.5) arc(-90:90:1.5 and 0.5);
      \coordinate[right=2cm of A7](r2);
      \draw[thick, shading=axis, right color=white, left color=custBlue] (r2)++(270:1.5 and 0.5) arc(270:90:1.5 and 0.5);
    \path[]
        (r1)++(35:1.5 and 0.5) edge [
          text=black,
          thick,
          shorten <=2pt,
          shorten >=2pt,
          bend left=50
          ] node[above]{\(\Gamma\)} (A1.north west);
    \path[]
        (r2)++(145:1.5 and 0.5) edge [
          text=black,
          thick,
          shorten <=2pt,
          shorten >=2pt,
          bend right=50
          ] node[above]{\(\Gamma\)} (A7.north east);

\end{tikzpicture}
\caption{(From Ref.~\onlinecite{Klockner_PRB_2020}.) Pictorial representation of the system discussed in Sec.~\ref{sec:fin_sys_res}: A tight-binding chain of $L$ sites with hoppings $t=1$ and interactions $U$ is driven out of equilibrium by zero-temperature reservoirs $\Gamma$ with different chemical potentials $\mu_\text{1,2}=\pm1$. }
\label{fig:sys_finite_chain}
\end{figure*}
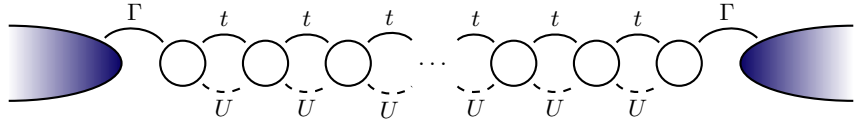%

\section{FINITE SYSTEMS: RESULTS}\label{sec:fin_sys_res}

We now apply the second-order FRG scheme to study a finite chain out of equilibrium. We follow Ref.~\onlinecite{Klockner_PRB_2020}, which contains additional details.

\subsection{Model: Finite tight-binding chain coupled to reservoirs at the edges}\label{subsubsec:model_fin}

We study a tight-binding chain with nearest neighbour hoppings $t$ and nearest neighbor interactions governed by (see Fig.~\ref{fig:sys_finite_chain}):
\begin{equation}\label{eq:ham_finite_chain}
\begin{split}
    H_\mathrm{sys}&=t\sum_{j=1}^{L-1} c_j^\dag c^\vdag_{j+1} +\mathrm{H.c.}+U\sum_{j=1}^{L-1} \left(c_j^\dag c^\vdag_j -\frac{1}{2}\right)\left(c_{j+1}^\dag c^\vdag_{j+1} -\frac{1}{2}\right).
    \end{split}
\end{equation}
In what follows, we always set $t=1$. We couple the system to physical left and right wide-band reservoirs labeled by $\nu=1,2$, respectively [see Eq.~\eqref{eq:wide_band_limit}]:
\begin{equation}\label{eq:physical_reservoirs}
\begin{split}
    \Gamma^{1,\mathrm{ret}}_{ij}&= -\I\Gamma \delta_{i1}\delta_{j1},~~~~\Gamma^{2,\mathrm{ret}}_{ij}=-\I\Gamma\delta_{iL}\delta_{jL}.
\end{split}
\end{equation}
These reservoirs are assumed to be at zero temperature $T_{\nu=1,2}=0$ and governed by different chemical potentials $\mu_{1}=-\mu_{2}=1$. 

All auxiliary reservoirs are identical and are characterized by a single temperature $T_\tn{cut}$ and a single chemical potential $\mu_\tn{cut}$. Different situations will be explored: i) $T_\tn{cut}=\mu_\tn{cut}=0$, ii) $T_\tn{cut}=0,\mu_\tn{cut}=\mu_\tn{2}=-1$, and iii) $T_\tn{cut}=\infty$, where in the last case, the choice of $\mu_{\tn{cut}}$ is irrelevant.

\begin{figure}[t!]
    \includegraphics[width=0.45\columnwidth]{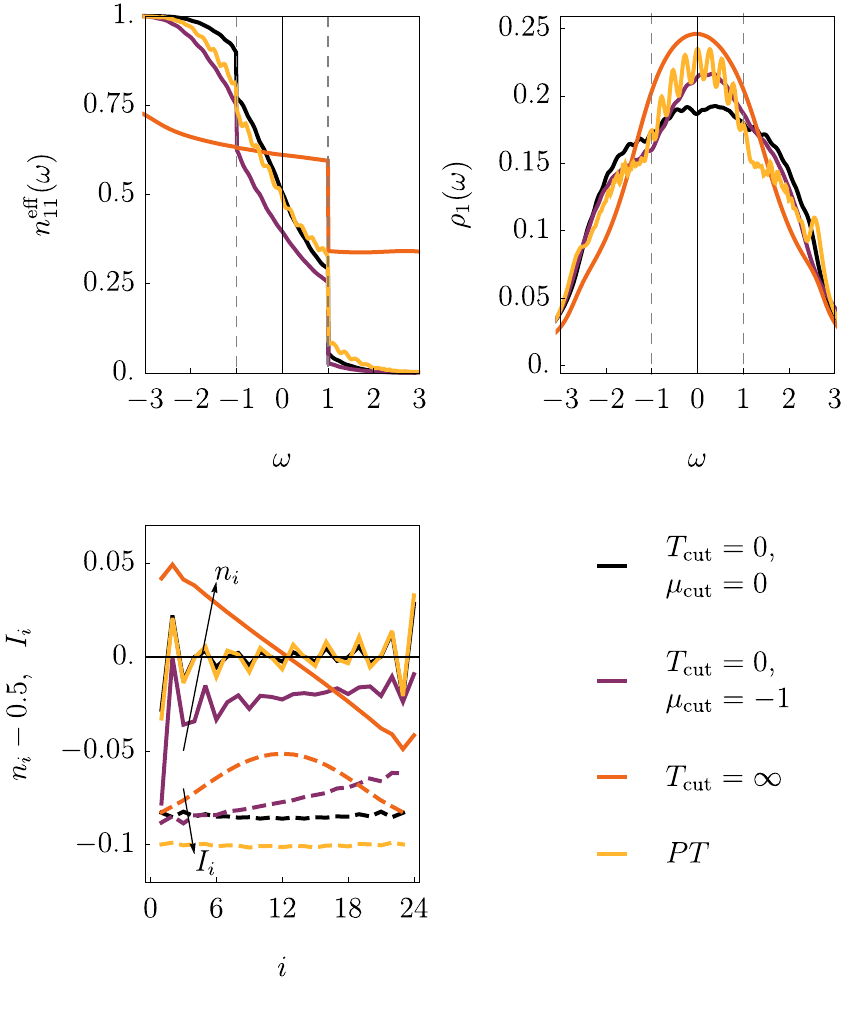}\hspace*{0.05\linewidth}
    \includegraphics[width=0.45\columnwidth,clip]{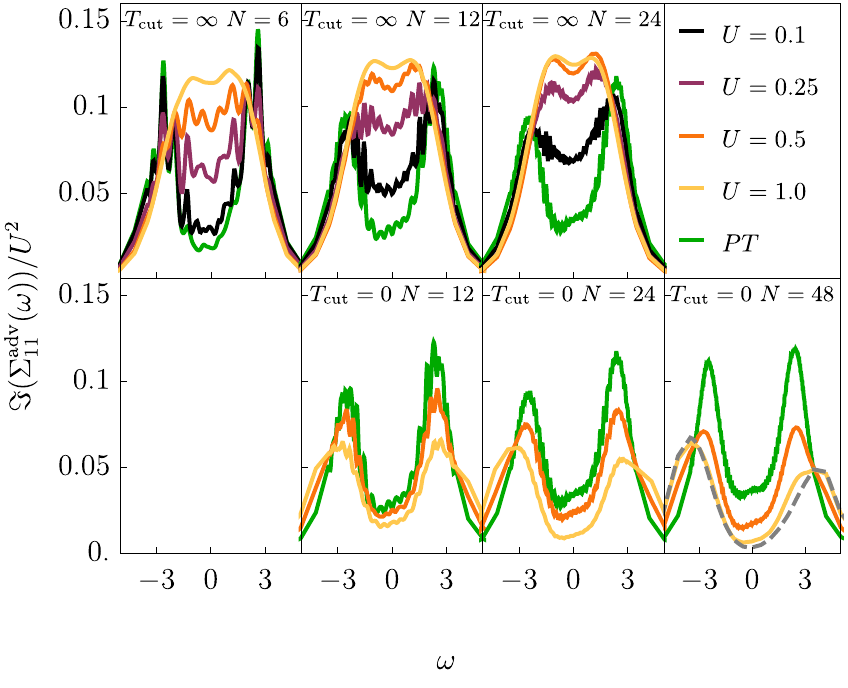}
    \caption{(Adapted from Ref.~\onlinecite{Klockner_PRB_2020}.) A chain of length $L=N$ with $t=1$ and \(\Gamma=0.2\) is driven out of equilibrium by a bias voltage $\mu_\text{1,2}=\pm1$. We compare FRG results obtained using various parameters of the auxiliary cutoff; the second-order perturbation-theory results are shown for comparison (PT). \emph{Left panels:} Fixed interaction $U=1$ and system size $L=N=24$. The plot shows the effective distribution function of Eq.~\eqref{eq:keldysh_sylvester_eq} at the boundary (top left), the occupation of the individual sites as well as the local current (solid and dashed; bottom left), and the local density of states $\rho_1(\omega)$ at the boundary (top right). \emph{Right panels:} Imaginary part of the self-energy at the boundary for various values of $U$ and $N=L$ at $\mu_\text{cut}=0$.
    }
    \label{fig:prb}
\end{figure}

\subsection{Fundamental questions to address}
\label{subsubsec:questions_fin}

The one-dimensional chain in Fig.~\ref{fig:sys_finite_chain} is governed by Luttinger liquid physics at low energies, and the Matsubara FRG was used to address the corresponding power-law behavior \cite{Meden_2003,Andergassen2004,Sbierski2017}. The fate of the Luttinger liquid phenomenology in non-equilibrium is being debated; e.g., a first-order FRG approach yields unconventional power-law exponents \cite{Jakobs2007}. It is unclear whether or not this scenario survives when inelatic scattering is accounted for by a second-order scheme. The framework of Sec.~\ref{sec:app_finite} can be employed to address this question; more details can be found in Ref.~\onlinecite{Klockner_PRB_2020}.

FRG results can be obtained for chains of up to $L=48$ sites using massive parallelization on hundreds of compute nodes \cite{Klockner_PRB_2020}. In equilibrium, the physical temperature and chemical potential provide a natural choice for the auxiliary reservoirs. In non-equilibrium, it is essential to check that the values of $T_\text{cut}$ and $\mu_\text{cut}$ do not influence the results. We will now demonstrate that this is not the case and that physical quantities qualitatively depend on the cutoff scheme. The FRG approach of Sec.~\ref{sec:app_finite} is hence inadequate to tackle the fate of non-equilibrium Luttinger liquid power laws \cite{Jakobs2007} in the presence of inelastic scattering.

\subsection{Application to the steady state of finite quantum chains out of equilibrium}

Fig.~\ref{fig:prb} shows FRG results for different temperatures $T_\text{cut}$ and chemical potentials $\mu_\text{cut}$ of the auxiliary reservoirs. We again stress that in contrast to equilibrium, there is no natural choice for these parameters. We compute the effective distribution function of Eq.~\eqref{eq:keldysh_sylvester_eq}, where one expects a piece-wise constant function with two steps in the limit of small $U$ and $\Gamma$. Moreover, we show the local occupation number $n_i=\left\langle c_i^\dagger c_i\right\rangle$, the particle currents $I_i$, and the local density of states at the edge
\begin{equation}
    \rho_i(\omega)=-\frac{1}{\pi}\text{Im}\left\{ \frac{1}{\left[ G^\mathrm{ret}(\omega)\right]^{-1} +\I \Gamma_\mathrm{smear}} \right\}_{ii},\ \Gamma_\mathrm{smear}=0.2.
\end{equation}
All of these quantities qualitatively depend on the FRG cutoff. This can be traced back to the behaviour of self-energy; the right panels of Fig.~\ref{fig:prb} illustrates the strong cutoff-dependence and the emergence of secular terms
\begin{eqnarray}
\text{Im\,}\left(\Sigma_{11}^{\text{adv}}(\omega)\right)/U^{2}\propto \text{pert.~theory in $U^2$} +  UN ,
\end{eqnarray}
which are only partly included in our approach. This illustrates that the current FRG scheme cannot adequately treat the system in Fig.~\ref{fig:sys_finite_chain}. More details can be found in Ref. \onlinecite{Klockner_PRB_2020}

\section{Summary}\label{sec:summary}

In this work, we reviewed and gave a detailed account of recently-developed, second-order FRG approaches to the steady-state of out-of-equilibrium problems in one dimension. The original works can be found in Refs.~\onlinecite{Klockner_NJP, Klockner_PRB_2020, Klockner_PRL, Klockner_PhD_thesis}. In particular, we discussed second-order truncation schemes which account for the flow of the two-particle vertex and which thus incorporate inelastic scattering (this is particularly important in non-equilibrium). The only fundamental approximation is the channel decomposition within the vertex flow equation. After introducing this general, common framework, we presented details on two different specific setups.

First, we recapitulated the application of the second-order FRG to the steady state of infinite systems which are translation invariant up to a shift in energy (Refs.~\onlinecite{Klockner_NJP,Klockner_PRL}). As an additional approximation, one assumes that the spatial range of the vertex functions is limited by a parameter $M$; the original flow equations are recovered in the limit $M\to\infty$, and convergence w.r.t.~$M$ can be reached in practice due to the translation invariance. The algorithm was used to study the non-equilibrium phase diagram of a generalized Wannier-Stark ladder coupled to reservoirs. One observes re-entrance behavior and multiple transitions between metallic and CDW phases. The FRG correctly captures the phase transition in equilibrium as well as in the mean-field limit of large $U$, which provides a highly non-trivial benchmark. Such a setup cannot be tackled straightforwardly using more accurate methods such as tensor networks and is a prototypical candidate for a FRG analysis.

Secondly, we reviewed the application of the second-order FRG to the steady-state of finite chains coupled to left and right reservoirs (Ref.~\onlinecite{Klockner_PRB_2020}). This setup is more involved due to the absence of translation invariance, and several additional approximations are needed to solve the FRG flow equations for the two-particle vertex in practice. In particular, we neglected the feedback of both the self-energy and the two-particle vertex in the flow equation of the two-particle vertex (the self-energy flow equation was not modified). This can be viewed as $\Lambda$-dependent perturbation theory and allows for an semi-analytic solution. Using large-scale parallelization, one can access systems of up to $\sim50$ sites. As a prototypical setup, we investigated an interacting tight-binding chain that is coupled to left and right reservoirs with different chemical potentials. It turned out, however, that this system cannot be treated reliably using the FRG scheme at hand since physical quantities depend qualitatively on the choice of the cutoff scheme.

While the second-order Keldysh FRG scheme provides a powerful tool to study the non-equilibrium phases of the generalized Wannier-Stark ladder, the method fails for finite systems. In this case, one needs to develop more involved truncation schemes that go beyond a purely perturbative treatment of the two-particle vertex. One potential way forward would be to shift the focus away from Luttinger liquid physics and to study small systems where the flow of the two-particle can be accounted for without having to resort to the drastic approximations used here. This was already done for impurity problems, which can serve as a natural starting point \cite{Hedden2004,Karrasch2008,Jakobs2010,Kinza2013,Laakso2014,Rentrop2014}. Another interesting avenue is a possible extension of the merger of dynamical mean field theory and FRG (DMF$^2$RG) \cite{PhysRevLett.112.196402}, which is currently used only in the equilibrium case, to non-equilibrium.

\section*{Acknowledgements}

G.C. and C.K. acknowledge support by the Deutsche Forschungsgemeinschaft (DFG, German Research Foundation) through the Emmy Noether program (KA3360/2-1) as well as by `Nieders\"achsisches Vorab' through the `Quantum- and Nano-Metrology (QUANOMET)' initiative within the project P-1. D.M.K. acknowledges funding by the Deutsche Forschungsgemeinschaft (DFG, German Research Foundation) under Germany's Excellence Strategy - Cluster of Excellence Matter and Light for Quantum Computing (ML4Q) EXC 2004/1 - 390534769 and support from the Max Planck-New York City Center for Non-Equilibrium Quantum Phenomena.

\section*{Author contributions}
All authors prepared the manuscript.

\section*{Data availability statement}
Data sharing not applicable to this article as no datasets were generated or analysed during the current study.

\bibliography{ref.bib}
 
\end{document}